# Data-driven Uncertainty Quantification in Computational Human Head Models

Kshitiz Upadhyay[1,2*], Dimitris G. Giovanis[3], Ahmed Alshareef[4], Andrew K. Knutsen[5], Curtis L. Johnson[6], Aaron Carass[4], Philip V. Bayly[7], Michael D. Shields[3], K.T. Ramesh[1,2]

[1]Hopkins Extreme Materials Institute, Johns Hopkins University, Baltimore, MD 21218, USA
[2]Department of Mechanical Engineering, Johns Hopkins University, Baltimore, MD 21218, USA
[3]Department of Civil and Systems Engineering, Johns Hopkins University, Baltimore, MD 21218, USA
[4]Department of Electrical and Computer Engineering, Johns Hopkins University, Baltimore, MD 21218, USA
[5]Center for Neuroscience and Regenerative Medicine, The Henry M. Jackson Foundation for the Advancement of Military Medicine, Bethesda, MD 20814, USA
[6]Department of Biomedical Engineering, University of Delaware, Newark, DE 19716, USA
[7]Mechanical Engineering and Materials Science, Washington University in St. Louis, St. Louis, MO 63130, USA

## Abstract

Computational models of the human head are promising tools for estimating the impact-induced response of the brain, and thus play an important role in the prediction of traumatic brain injury. The basic constituents of these models (i.e., model geometry, material properties, and boundary conditions) are often associated with significant uncertainty and variability. As a result, uncertainty quantification (UQ), which involves quantification of the effect of this uncertainty and variability on the simulated response, becomes critical to ensure reliability of model predictions. Modern biofidelic head model simulations are associated with very high computational cost and high-dimensional inputs and outputs, which limits the applicability of traditional UQ methods on these systems. In this study, a two-stage, data-driven manifold learning-based framework is proposed for UQ of computational head models. This framework is demonstrated on a 2D subject-specific head model, where the goal is to quantify uncertainty in the simulated strain fields (i.e., output), given variability in the material properties of different brain substructures (i.e., input). In the first stage, a data-driven method based on multi-dimensional Gaussian kernel-density estimation and diffusion maps is used to generate realizations of the input random vector directly from the available data. Computational simulations of a small number of realizations provide input-output pairs for training data-driven surrogate models in the second stage. The surrogate models employ nonlinear dimensionality reduction using Grassmannian diffusion maps, Gaussian process regression to create a low-cost mapping between the input random vector and the reduced solution space, and geometric harmonics models for mapping between the reduced space and the Grassmann manifold. It is demonstrated that the surrogate models provide highly accurate approximations of the computational model while significantly reducing the computational cost. Monte Carlo simulations of the surrogate models are used for uncertainty propagation. UQ of the strain fields highlights significant spatial variation in model uncertainty, and reveals key differences in uncertainty among commonly used strain-based brain injury predictor variables.

---

*Corresponding author: kshitiz@jhu.edu (Kshitiz Upadhyay)

## Keywords

## 1. Introduction

Traumatic brain injury (TBI) is one of the leading causes of mortality and morbidity in the world, with the latest data from United States showing nearly 61,000 TBI-related fatalities in 2019 [1]. Typically caused by the rapid application of external forces to the head, TBI can lead to a host of disabilities: lost or impaired consciousness, memory loss, confusion, disorientation, altered vision, etc. [2,3]. Given the mechanical origins of TBI, biofidelic computational head models have been extensively used to study the deformation of the human brain within the head under rapid loading conditions; such deformations have been correlated to increased risk of brain injury [4–6]. Computational head models are thus playing a critical role in bridging the gap between external mechanical insult to the head and the resulting neuropathology.

Within TBI research, a computational head model has three primary components: the head geometry (based on the anatomy), the material properties of the various tissues, and the boundary conditions [7]. Each of these primary components is associated with considerable variability. For example, a study [8] on the brain volume of 52 healthy humans (both males and females) found a size difference of ~81% between the largest and the smallest brains. Similar differences are also seen in the case of the material properties. For example, between three experimental studies [9–11] in the literature, the reported long-term shear modulus of brain tissue varies between 12.6 Pa to 303.3 Pa. A considerable variation in material property values within different brain regions is also reported by several in-vivo experimental studies [12,13]. Recent literature has shown that such variability in head shape/size and brain tissue material properties can result in very different predicted brain deformations from computational head models [14,15]. Such discrepancies in reported strains can lead to very different injury risk predictions, which makes it more difficult to apply these head models in real-world injury scenarios. Despite this, a majority of the available human head models are developed using deterministic inputs of head geometry (e.g., from a 50th-percentile male [16]), material properties, and boundary conditions. Uncertainty quantification (UQ) of head models, which involves quantification of the effect of variability and uncertainty in the input parameters on the model output (e.g., strain fields), has received little attention.

UQ of computational head models poses significant challenges. For instance, a typical forward uncertainty propagation requires running a large number of simulations with different realizations of the input parameters (as a random vector). However, high fidelity computational head model simulations, which feature nonlinear constitutive models and complex geometrical features, generally run for hours to days [17,18], and so their UQ can become prohibitively expensive. Surrogate models can alleviate this issue to an extent by providing a low-cost approximation of the computational model. However, creating accurate surrogate models with the small number of available training data due to limited number of feasible computational model runs is non-trivial, and requires careful sampling of realizations from the input parameter space. Further, complex biofidelic head models are associated with high dimensional inputs and outputs; for example, a typical finite element head simulation results in a strain field at tens of thousands of nodes [18]. Such a high dimensionality of model inputs and outputs can lead to poor accuracy of the surrogate model and issues such as the curse of dimensionality and over-fitting [19,20]. Due to these challenges, existing studies on the UQ of head models in the brain biomechanics community have only focused on low-fidelity, low-dimensional head models [21–24].

In this study, a data-driven manifold learning-based surrogate modeling framework is proposed for UQ of high-fidelity, high-dimensional computational human head models. Although the proposed framework is applicable to a wide variety of cases where variability or uncertainty in input parameters leads to uncertainty in model outputs (e.g., variability in the head shape across the human population), a specific model problem is chosen for the purposes of this work: 2D subject-specific human head model of a 31-year-old male subject. In this model, the anatomically accurate head geometry is derived from magnetic resonance imaging (MRI) data, while the nonlinear visco-hyperelastic material parameters of different brain regions are derived using magnetic resonance elastography (MRE). While the head geometry and boundary conditions (mild rotational acceleration of the head) are assumed to be deterministic, the material parameters within individual brain regions can vary significantly. The objective is to study the effect of this variability on two time-independent scalar strain fields: the maximum axonal strain (MAS) and the cumulative maximum principal strain (CMPS). In the first stage of the proposed UQ framework, the probabilistic sampling method introduced in [25] is utilized to generate realizations of the high-dimensional (input) random vector containing the material properties (based on the available experimental MRE data). In the second stage, a surrogate model is trained to create a low-cost and accurate mapping between the material properties and the MAS/CMPS fields. The surrogate model employs manifold learning (specifically, Grassmannian diffusion maps [26]) to identify a latent representation of the MAS/CMPS fields, and machine learning (i.e., Gaussian process [27] and geometric harmonics [28]) to create a mapping between (i) the input random vector and the reduced solution space and, (ii) the reduced space and the Grassmann manifold. The proposed framework accelerates the UQ of the full MAS/CMPS fields, given the uncertainty in the head model material properties.

The paper is organized as follows: Section 2 describes the 2D subject-specific computational head model and its associated uncertainties. In Section 3, the proposed data-driven UQ framework is formalized, and the surrogate model is trained to create a mapping between the MRE material properties and the full-field strain maps. Section 4 discusses the performance of the proposed surrogate in predicting the full strain fields. The uncertainty in the predicted strain fields and in several scalar strain measures commonly employed for the quantification of brain injury, are reported.

## 2. Subject-Specific 2D Head Model

In a recent study by the authors [29], a subject-specific 3D head model was developed for a healthy 31-year-old male subject (illustrated in Fig. 1), using head geometry derived from magnetic resonance imaging (MRI) and visco-hyperelastic material properties calibrated from in-vivo magnetic resonance elastography (MRE) [30,31]. This model was employed to simulate brain deformation during mild rotational accelerations of the head about the inferior-superior (z-) axis, using a meshless, material point method (MPM) numerical technique. The simulation results were validated using experimentally observed time-varying strain fields (under the same loading conditions) acquired via tagged MRI (tMRI) of a neck rotation experiment [32], conducted on the same human subject. In this work, a 2D version of this validated 3D head model (see Fig. 1(g)) is selected to quantify uncertainty in the strain fields resulting from the simulation of neck rotation motion, given the uncertainty in the input material properties. Note that 2D head simulations of such loading conditions are common in the literature because of the nearly planar deformation fields (i.e., negligible out-of-plane motion in the z-direction) [33,34].

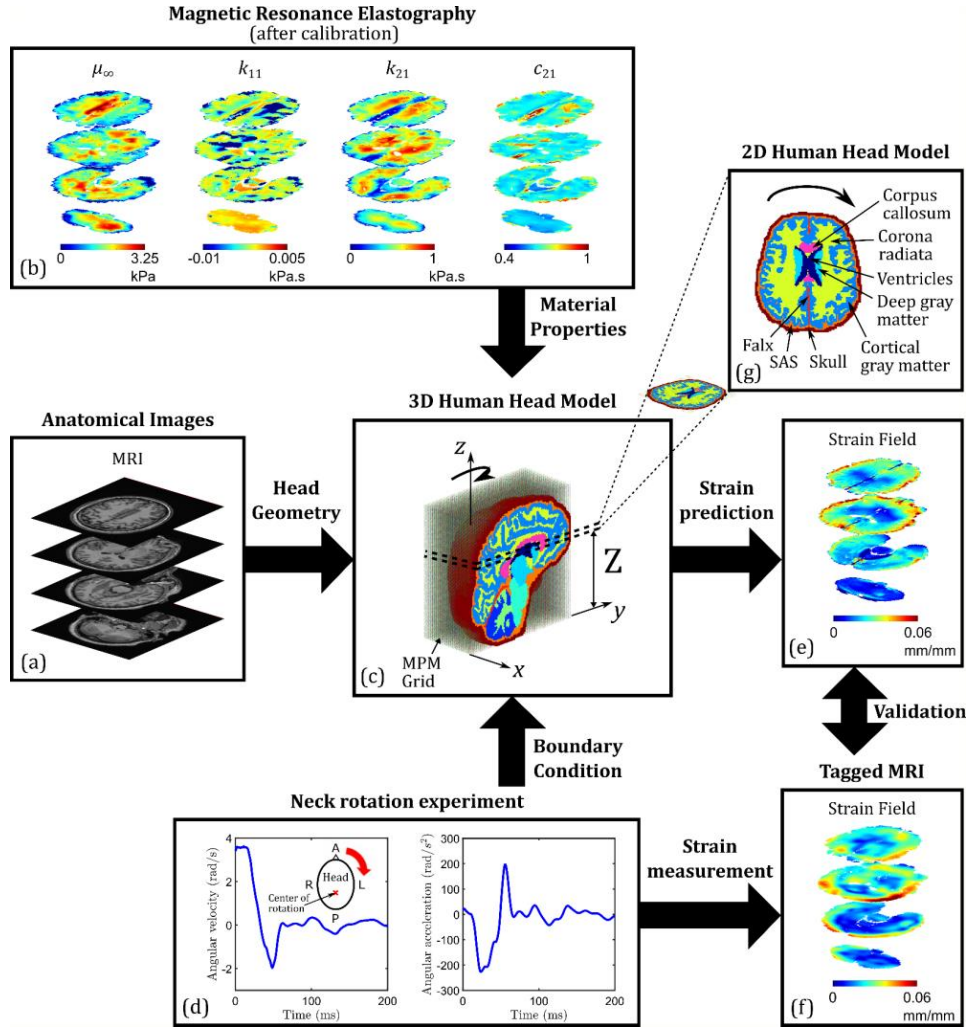

**Figure 1**. Flowchart of the development of subject-specific computational human head models. (a) Segmented and processed anatomical images from MRI (including diffusion tensor imaging) provide the 3D head geometry and axonal fiber orientation, while (b) in-vivo MRE is used to calibrate nonlinear visco-hyperelastic material properties of major brain substructures; (c) the resulting MPM-based 3D head model is used to simulate a (d) neck rotation experiment. (f) Tagged MRI of this experiment yields full-field strain data, which is compared with the (e) simulated strain data for model validation. In this study, a (g) 2D slice of the 3D head model is considered for UQ purposes.

Critical aspects of the subject-specific 2D head model will now be discussed in greater detail. The first subsection describes the measured (using tMRI) full-field strain response, which dictates the choice of the specific plane for 2D model development (i.e., the axial layer defined by the distance Z in Fig. 1(c)). The second subsection briefly discusses the geometry and brain morphometry of the 2D head model along with the constitutive modeling and numerical simulation frameworks (detailed descriptions are available in the original 3D model article [29]). Finally, the last subsection describes the uncertainties associated with the head model, which serve as a motivation for this work.

## 2.1. Tagged MRI and the choice of axial brain layer

As mentioned before, tMRI was employed in Upadhyay et al. [29] to obtain full-field 3D displacements and strain fields from neck rotation experiments (see experimental details in [7,32]) on a 31-year-old human subject (Fig. 1(d),(f)). In this experiment, a controlled non-injurious impulsive loading is applied on the subject's head, which rotates in the axial plane about the inferior-superior (I/S) axis (center of rotation roughly passes through the brain stem). The loading input to the head is measured using an angular position sensor, which also provides boundary condition to the computational model. Figure 1(d) shows the angular velocity and acceleration versus time plots. Time-varying Green-Lagrange (G-L) strain tensor fields during this loading condition are measured at a spatial resolution of 1.5 mm (isotropic) and a temporal resolution of 18 ms, using tMRI. These tensorial strain fields are generally reduced to two scalar strains during post-processing [35]: (i) the maximum principal strain, MPS (first principal eigenvalue of the G-L strain tensor), and (ii) the axonal strain, $E_f$. The latter is computed as,

$$E_f = \boldsymbol{a_0} \cdot \mathbf{E} \cdot \boldsymbol{a_0} \tag{1}$$

where $\boldsymbol{a_0}$ is the axonal fiber direction at a given brain voxel obtained from diffusion tensor imaging (DTI), and $\mathbf{E}$ is the experimentally-derived G-L strain tensor. It is important to note that while MPS is computed for the entire brain volume, $E_f$ is only computed for brain voxels in the anisotropic white matter regions (i.e., corpus callosum, corona radiata, cerebellar white matter, and brainstem) with a fractional anisotropy (FA) value greater than or equal to 0.2 (the FA field is also obtained from DTI). This is a standard criterion [36] that is implemented to exclude regions of isotropic diffusion such as gray matter and cerebrospinal fluid.

The strains in the brain tissue are correlated to injuries such as diffuse axonal injury (DAI) [32]: higher values of scalar strains at a given location in the brain are generally related to a greater probability of injury. As a result, the cumulative MPS (CMPS) and maximum axonal strain (MAS), which are the peak values of MPS and $E_f$ at a given brain voxel over the entire loading duration, respectively, are commonly used in metrics for the quantification of brain injury risk [4–6,37]. Note that both CMPS and MAS are time-independent scalar strains. In this study, the axial layer of the brain in which the greatest area fraction experiences MAS that is greater than the thresholds of 50- and 95-percentile MAS (evaluated over the entire 3D brain volume) is selected for 2D head model development. This brain layer is potentially the most vulnerable to injury under this loading condition.

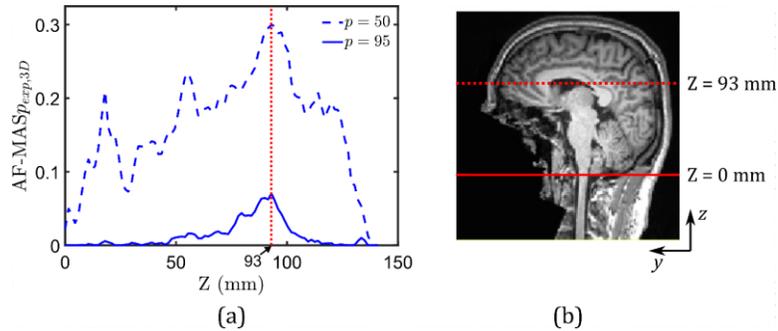

(a)          (b)

**Figure 2.** (a) Area fractions of the brain axial layers/slices experiencing MAS that is greater than the global 50- and 95-percentile MAS values, as a function of their position in the z-direction (along the Inferior-Superior axis). A generic MAS area fraction-based metric for $p^{\text{th}}$ percentile MAS threshold is denoted by AF-MAS$p_{\text{exp,3D}}$. (b) T1-weighted MRI image of the midsagittal ($yz$-) plane, showing the position of axial layers at Z = 0 and 93 mm.

This identification process is shown in Figure 2, where Fig. 2(a) plots the area fractions of different axial layers exposed to MAS greater than the 50- (dashed line) and 95-percentile (solid line) MAS thresholds as a function of their z-location. Note that Z = 0 mm, indicated on the midsagittal slice of the T1-weighted MRI image in Fig. 2(b), corresponds to the bottom-most brain voxel (excluding the sub-arachnoid space and the skull) in the 3D head model in Fig. 1(c). Both the 50- and 95-percentile MAS area fraction-based metrics reach their maxima at Z = 93 mm. Thus, this particular axial layer/slice, which passes through the genu of corpus callosum, is chosen for the 2D head model development in this work.

Finally, full-field 3D displacements observed from tMRI are also used to estimate the out-of-plane displacement of material points in the 2D head model: the observed average z-displacement of the Z = 93 mm layer from tMRI is compared with the corresponding x- and y-displacements. It is found that the average z-displacement is only ~10% of the average x- and y- displacements, which provides a reasonable justification for using a zero z-displacement constraint in the ensuing 2D model simulations.

## 2.2. Model development

In Upadhyay et al. [29], subject-specific 3D anatomical images of the subject head acquired using MRI were processed and segmented at a spatial resolution of 1.5 mm (isotropic) into thirteen smaller substructures: deep gray matter, cortical gray matter, corona radiata, corpus callosum, cerebellum gray matter, cerebellum white matter, brainstem, ventricles, cerebrospinal fluid (CSF), falx, tentorium, sub-arachnoid space (SAS), and the skull. In this study, a single axial slice of this 3D geometry (Fig. 1(c)), corresponding to Z = 93 mm, is used as the 2D head model geometry for neck rotation simulation. Note that data from all the biomedical imaging techniques (MRI (including DTI), MRE and tMRI) was co-registered to a common coordinate space, allowing a one-to-one correspondence of brain voxel locations between geometry, material properties, and experimental strain fields. There are eight substructures present in the 2D head model (see Fig. 1(g)): deep gray matter, cortical gray matter, corona radiata, corpus callosum, ventricles, falx, SAS, and skull.

In-vivo MRE was conducted on the same human subject to acquire spatially-resolved, full-field maps of the shear storage and loss moduli at three actuation frequencies (i.e., 30, 50, and 70 Hz). Due to its coarse resolution (i.e., 1.5 mm isotropic), the MRE maps only consist of the four major parenchymal brain substructures: deep gray matter, cortical gray matter, corona radiata, corpus callosum. The frequency-dependent storage and loss moduli at every MRE brain voxel are combined with stress-strain data from ex-vivo experiments on human brain tissues to calibrate (see details in Upadhyay et al. [29]) a nonlinear visco-hyperelastic Ogden-Upadhyay-Subhash-Spearot (O-USS) constitutive model [38]. A brief description of the O-USS model is provided in the supplementary material. It consists of six model parameters: $\mu_\infty$ — the long-term shear modulus, $\alpha$ — the compression-tension asymmetry parameter, $k_{11}$ — the linear rate sensitivity control parameter, $k_{21}$ — the nonlinear rate sensitivity control parameter, $c_{11}$ — the rate sensitivity index, and $\kappa$ — the bulk modulus.

The calibration of full-field visco-hyperelastic material properties from MRE and the subsequent consideration of a single axial layer at Z = 93 mm yields four O-USS model parameters, viz., $\mu_\infty$, $k_{11}$, $k_{21}$, and $c_{11}$, for every voxel in the deep gray matter, cortical gray matter, corona radiata, and corpus callosum substructures of the 2D head model. Note that the parameter $\alpha$, which captures the nonlinearity in the quasi-static stress-strain response under large deformations, cannot be calibrated from in-vivo MRE because MRE probes the material only in its small deformation regime. Similarly, bulk modulus $\kappa$ also cannot be obtained from MRE, which assumes incompressibility of material response. Therefore, while $\alpha$ is directly calibrated for each of the four major substructures from their ex-vivo stress versus strain responses from the literature [39,40], a constant value of $\kappa$ for brain tissue is taken from the literature [41,42]. Finally, the material properties of the finer brain substructures that were unresolved in MRE (i.e., SAS, falx, skull, and ventricles) are taken directly from ex-vivo experimental data in the literature [41,43–45]. Among these minor regions, the SAS is modeled as a soft linear viscoelastic (LVE) solid characterized by a short-term shear modulus $G_0$, fitting parameter $g_1$, and time-constant $\tau_1$, while the falx and skull are modeled as linear elastic solids characterized by a Young's modulus $E$ and a Poisson's ratio $\nu$. Ventricles are modeled as a viscous fluid using shear viscosity $\mu$ and the parameter $n$ of the Murnaghan-Tait equation of state [46]. For more details on these constitutive formulations, refer the original 3D model development article [29]. The average (mean) material properties of all the eight substructures of the 2D head model are listed in Table 1.

**Table 1**. Average material properties of the various substructures of the 2D head model employed in this work.

| Brain substructure | Material properties (O-USS model) | | | | | | |
|---|---|---|---|---|---|---|---|
| | $\mu_\infty$ (kPa) | $\alpha$ | $k_{11}$ (kPa.s) | $k_{21}$ (kPa.s) | $c_{21}$ | $\kappa$ (GPa) | $\rho$ (kg/m³) |
| Deep gray matter | 1.10 | 4.92 | -0.152 | 0.584 | 0.879 | | |
| Cortical gray matter | 1.35 | -3.76 | -0.461E-2 | 0.441 | 0.672 | 2.19 | 1040 |
| Corpus Callosum | 2.11 | -2.32 | -0.224E-2 | 0.517 | 0.598 | | |
| Corona Radiata | 1.73 | -3.47 | -0.343E-2 | 0.635 | 0.635 | | |

| | Material properties (LVE model) | | | | | |
|---|---|---|---|---|---|---|
| | $G_0$ (kPa) | $g_1$ | $\tau_1$ (ms) | $\kappa$ (GPa) | $\rho$ (kg/m³) | Reference |
| Sub-arachnoid space | 0.5 | 0.8 | 12.5 | 2.19 | 1133 | Mao et al. (2013) [41] |

| | Material properties (Linear elastic model) | | | |
|---|---|---|---|---|
| | $E$ (MPa) | $\nu$ | $\rho$ (kg/m³) | Reference |
| Falx | 31.5 | 0.45 | 1133 | Galford and McElhaney (1970) [43] |
| Skull | 8000 | 0.22 | 2070 | McElhaney (1973) [44] |

| | Material properties (Fluid model) | | | |
|---|---|---|---|---|
| | $\mu$ (Pa.s) | $n$ | $\rho$ (kg/m³) | Reference |
| Ventricles | 1.002e-3 | 7.15 | 1004 | Goldsmith (1970) [45] |

Finally, the 2D subject-specific head model is used to simulate mild rotational acceleration (Fig. 1(d)) of the head over a 189 ms duration, using the Uintah software MPM package, as described previously [29]. The simulation results in the time-varying full-field G-L strain maps of the 2D brain at a 1.5 mm spatial resolution and a 3 ms temporal resolution. The tensorial G-L strain is used to compute the time-independent scalar strain fields of MAS and CMPS during post-processing. The comparison of the simulated strain-response from the head model employing mean material properties (Table 1) with corresponding observed strain-response from tMRI, is presented in the supplementary material (Figure S1). A reasonable agreement is observed both in terms of the magnitudes (evaluated at 95-percentile) of strains $\mathbf{E}$ (note, the tensor $\mathbf{E}$ has three in-plane components: $E_{xx}$, $E_{xy}$, and $E_{yy}$), $E_f$, and MPS, and their evolution in time. Thus, the 2D subject-specific head model considered in this study is a validated model.

## 2.3. Sources of uncertainty in the 2D head model

The process of identifying sources of uncertainty in a computational model is a complex task. Based on the classification provided in [47], Iliopoulos et al. [21] listed the following categories of uncertainty sources that can be present in a computational model of the human head:

- Parameter variability — introduced by variability in the input parameters of the model (e.g., variability in head geometry across human population, or variability in material properties of the brain substructures for a particular human subject).
- Parameter uncertainty — introduced by unknown model parameters whose values are estimated from experiments or statistical methods (e.g., uncertainty from calibration of constitutive model parameters).
- Model inadequacy — introduced by lack of knowledge of underlying true physics or by approximations in the modeling approach to capture the known physics (e.g., uncertainty due to constitutive model approximations, model retaining only lower-order terms, etc.).
- Numerical uncertainty — introduced by numerical issues of the computational model (e.g., uncertainty from numerical errors in MPM simulations).
- Observation errors — introduced by uncertainty in experimental measurements (e.g., uncertainty in MRE shear moduli measurement).
- Interpolation uncertainty — introduced by lack of available data in the model's range of applicability (e.g., uncertainty coming from the application of a constitutive model, which was calibrated from stress-strain data at only a few loading rate values, to predict material response in a continuous strain rate spectrum during simulations).

The UQ framework developed in this work (see Section 3) can quantify the effect of parameter variability and parameter uncertainty on the model output, when input parameters (i.e., the model geometry, material properties, and boundary conditions) can be represented via probability distributions. Other sources of uncertainty — model inadequacy, numerical uncertainty, observation errors, and interpolation uncertainty — are not explicitly considered.

Specifically, for the model problem of the subject-specific 2D computational head model (Fig. 1) considered in this study, both the head geometry, which is derived directly from anatomical images of the subject's head, and boundary condition, which is directly measured using an angular position sensor during the neck rotation experiment, are considered to be deterministic. However, the material properties that are assigned to each of the brain substructures constitute a significant source of uncertainty, which affects the model outputs (e.g., simulated strain fields). Figure 3(a) shows box plots of the four visco-hyperelastic material properties derived from in-vivo MRE for each of the four major substructures of the 2D model (note, each substructure is composed of a number of brain voxels, each with one set of material properties).

Significant inter-region variability of these material properties is evident from these plots; despite this, the computational model assumes homogeneous material property values within individual regions (i.e., a single set of properties is assigned to each brain substructure during simulations) and does not account for spatial variability of material properties, which leads to uncertainty in the simulated response. To highlight the effect of this parameter variability on the simulated response, Fig. 3(b) compares the MAS and CMPS strain fields of the head model for two different sets of material property values: mean (Table 1) and a randomly sampled set ("sample set"), assigned to each of the four brain regions (both sets are highlighted on the box plots). Figure 3(b) shows that in the case of the randomly sampled material properties, the simulation predicts larger overall strain values compared to the case using the mean material properties. Quantitatively, the 50- and 95-percentile MAS of the former simulation are 0.011 mm/mm and 0.037 mm/mm, respectively, which is 66.67% and 12.12% higher compared to the mean material properties case; similarly, the 50- and 95-percentile CMPS are 14.85% and 17.10% higher. The goal of the present study is to quantify uncertainty in the strain outputs of the subject-specific 2D head model originating from the variability of material properties within individual brain substructures. Specifically, the MAS and CMPS strain field outputs are considered.

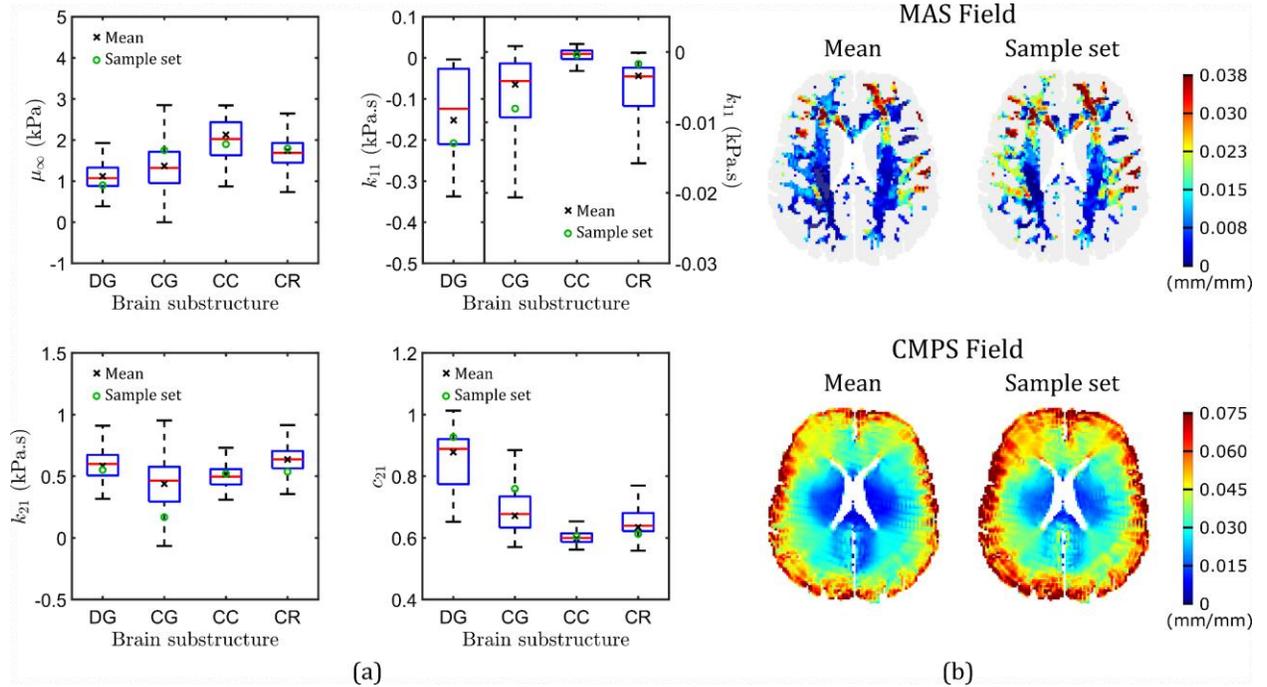

(a)

(b)

**Figure 3**. (a) Box plots of the four visco-hyperelastic material properties calibrated for brain voxels in the four major brain substructures: deep gray matter (DG), cortical gray matter (CG), corpus callosum (CC), and corona radiata (CR). On each of the boxes, the central red line indicates the median, and the bottom and top edges indicate the 25th and 75th percentiles, respectively. Maximum whisker length is 1.5 times the interquartile range. Two sets of material property values are also indicated: mean material properties, and a randomly sampled set of properties ("Sample set"). (b) Comparison of MAS and CMPS strain fields simulated by two head models: one based on mean material properties, and the other based on the "Sample set".

## 3. **Proposed Uncertainty Quantification framework**

## 3.1. Problem statement

Consider a complete probability space $[\Omega, F, P]$, where $\Omega$ is the sample space, $F \subset 2^{\Omega}$ is the $\sigma$-algebra of the events, and $P$ is the probability measure. Furthermore, let $X(\omega) = \{X_1(\omega_1), \dots, X_n(\omega_n)\} \in X \subset \mathbb{R}^n$ denote a vector of $n$ random variables having probability density function (pdf) $f_X(x)$, indexed on $\omega = \{\omega_1, \dots, \omega_n\} \in \Omega = [0, 1]^n$. The stochastic formulation of the present problem is defined as

$$\mathcal{M}^{(\text{MAS,CMPS})}\big(X_{\text{M}}(\omega)\big) = Y_{\text{M}}^{(\text{MAS,CMPS})} \tag{2}$$

where $\mathcal{M}(\cdot)$ is the 2D head model, $X_{\text{M}}(\omega)$ is the vector containing the material parameters, and $\mathcal{M}^{\text{MAS}}$ and $\mathcal{M}^{\text{CMPS}}$ represent mappings to the random vectors of the output MAS and CMPS fields, $Y_{\text{M}}^{\text{MAS}}$ and $Y_{\text{M}}^{\text{CMPS}}$, respectively. For the 2D subject-specific head model, vector $X_{\text{M}} \in \mathbb{R}^{16}$ ($\omega$ is omitted for the remaining of the paper) consists of the four visco-hyperelastic material properties (i.e., $\mu_\infty$, $k_{11}$, $k_{21}$, and $c_{21}$) for all four brain substructures (i.e., deep gray matter (superscript DG), cortical gray matter (superscript CG), corona radiata (superscript CR), and corpus callosum (superscript CC)), derived from in-vivo MRE:

$$X_{\text{M}} = \big(\mu_\infty^{\text{DG}}, k_{11}^{\text{DG}}, k_{21}^{\text{DG}}, c_{21}^{\text{DG}}, \mu_\infty^{\text{CG}}, k_{11}^{\text{CG}}, k_{21}^{\text{CG}}, c_{21}^{\text{CG}}, \mu_\infty^{\text{CR}}, k_{11}^{\text{CR}}, k_{21}^{\text{CR}}, c_{21}^{\text{CR}}, \mu_\infty^{\text{CC}}, k_{11}^{\text{CC}}, k_{21}^{\text{CC}}, c_{21}^{\text{CC}}\big) \in \mathbb{R}^{16} \tag{3}$$

The output random vector $Y_{\text{M}}^{\text{MAS}} \in \mathbb{R}^{2125}$ consists of MAS values at brain voxels in the structurally anisotropic white matter regions (i.e., the corona radiata and corpus callosum). On the other hand, $Y_{\text{M}}^{\text{CMPS}} \in \mathbb{R}^{6372}$ consists of brain voxels in all four brain substructures, and thus is a larger vector compared to $Y_{\text{M}}^{\text{MAS}}$. Every component in these two vectors corresponds to a specific voxel in the 2D brain.

To perform UQ, it is necessary to draw samples from the joint pdf $f_{X_{\text{M}}}(x)$ of $X_{\text{M}}$ and run the computational model $\mathcal{M}(\cdot)$. However, this joint pdf is not known a priori. To this end, the data-driven methodology introduced in [25] is utilized in this study to sample realizations of $X_{\text{M}}$ that are statistically consistent with the available in-vivo MRE data. This process is described in Section 3.2. Once a large number of realizations of the input random vector $X_{\text{M}}$ are generated, uncertainty can be propagated through the model $\mathcal{M}(\cdot)$ to calculate the statistical properties of the MAS and CMPS strain field outputs, respectively. However, due to the excessive computational cost of running computational head models (for instance, a single 2D subject-specific head model simulation runs for several hours on a high performance computing (HPC) cluster), propagation of uncertainty directly via $\mathcal{M}(\cdot)$ is not feasible.

To overcome this bottleneck, the present study proposes development of a surrogate model $\widetilde{\mathcal{M}} \equiv \mathcal{M}(\cdot)$ that will reasonably approximate the strain response of the full computational head model (i.e., $\widetilde{\mathcal{M}}(X_{\text{M}}) \approx \mathcal{M}(X_{\text{M}})$) in a fraction of the computational time required by the model $\mathcal{M}(\cdot)$. To create the training data for the surrogate model, the computationally expensive 2D head model is run for a small number of $X_{\text{M}}$ realizations that span the probability space efficiently. Once trained, the surrogate will be used in the framework of Monte Carlo simulation to predict the MAS and CMPS fields and calculate their statistical properties. The details of the surrogate model are discussed in Section 3.3.

## 3.2. Data-driven sampling

The data-driven methodology introduced in [25] is employed to generate realizations of the random vector $X_{\text{M}}$ that are statistically consistent with the available in-vivo MRE data of the human subject on which the 2D subject-specific computational head model is based. Briefly, the method utilizes a multi-dimensional

Gaussian kernel-density estimation to obtain the probability distribution of the scaled and normalized data. Then, diffusion maps is used to reveal the local geometry of the subset $\mathcal{S} \subset \mathbb{R}^{16}$ on which the probability distribution is concentrated. Diffusion maps require choosing a Gaussian kernel smoothing parameter ($\varepsilon$) and a parameter $\kappa$ that is used to fix the analysis scale of the local geometric structure of the dataset (refer to [25] for details). Finally, Markov Chain Monte Carlo (MCMC) based on Itô stochastic differential equations is utilized to sample realizations of the random vector that are confined in $\mathcal{S}$. However, for the 2D head model, one challenge is that the in-vivo MRE data are, in some sense, heterogeneous; one set of material properties is available per brain voxel, but the number of voxels in the four brain substructures is different. To overcome this, the present study proposes a two-step strategy for generating data for the entire 2D head model that are consistent with the available in-vivo MRE data:

**Step 1:** For each one of the four individual substructures, generate realizations of the random vector $\boldsymbol{X}_{\mathrm{M}}^{i} \in \mathbb{R}^4$:

$$\boldsymbol{X}_{\mathrm{M}}^{i} = \left( \mu_{\infty}^{i}, k_{11}^{i}, k_{21}^{i}, c_{21}^{i} \right) \tag{4}$$

where $i \in \{ \mathrm{DG}, \mathrm{CG}, \mathrm{CR}, \mathrm{CC} \}$. In this case, the in-vivo MRE data in each one of the brain substructures, used to drive the sampling, consists of 300 points randomly selected from the material properties. 900 realizations are generated using the data-driven method, and 100 realizations (out of a total of 900) are randomly selected to represent substructure $i$. By definition, data $D_i \in \mathbb{R}^{4 \times 100}$ are statistically consistent with the in-vivo MRE data for the brain substructure $i$.

**Step 2:** Combine data $D_i$ from all brain substructures to create a dataset $D \in \mathbb{R}^{16 \times 100}$ to drive the generation of realizations of the random vector $\boldsymbol{X}_{\mathrm{M}} \in \mathbb{R}^{16}$ that contain the material properties for the entire 2D head model. Having identified dataset $D$ that is "implicitly" consistent with MRE data, 10,200 additional realizations of $\boldsymbol{X}_{\mathrm{M}}$ are then generated ($i^{\mathrm{th}}$ realization is denoted by $\boldsymbol{x}_{\mathrm{M}}^{(i)}$). Out of the total 10,300 realizations, 300 are used as the training set for the surrogate model (presented next), and 10,000 are used for performing Monte Carlo simulations using the trained surrogate.

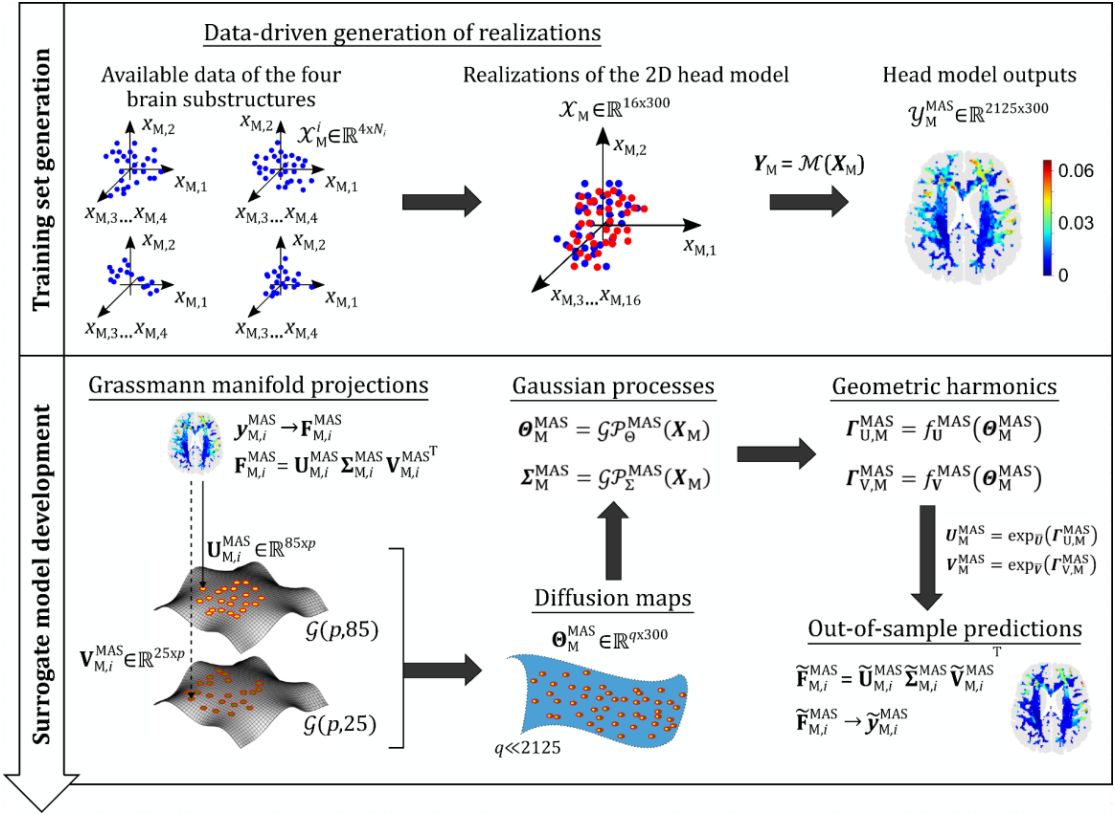

**Figure 4**. A schematic illustration of the proposed data-driven surrogate modeling framework for UQ of computational head models: in the first stage (see Section 3.2), the available material properties of each of the four substructures, $\mathcal{X}_{\mathrm{M}}^i \in \mathbb{R}^{4 \times N_i}$ ($N_i$ denotes number of voxels for substructure $i$) are used to generate 300 realizations of the input random vector of material properties for the 2D head model (i.e., $\mathcal{X}_{\mathrm{M}}$). Simulations of these realizations yields input-output ($\mathcal{X}_{\mathrm{M}} - \mathcal{Y}_{\mathrm{M}}^{\mathrm{MAS}}$) data sets for training the surrogate model in the second stage. The surrogate model is developed in three steps (see Section 3.3): 1. perform nonlinear dimensionality reduction on the output via Grassmannian diffusion maps, 2. create Gaussian process mappings between the input and the reduced solutions (i.e., $\Theta_{\mathrm{M}}^{\mathrm{MAS}}$ and $\Sigma_{\mathrm{M}}^{\mathrm{MAS}}$), and 3. for out-of-sample predictions, create geometric harmonics mappings between the diffusion coordinates $\Theta_{\mathrm{M}}^{\mathrm{MAS}}$ and the matrices $\Gamma_{\mathrm{U,M}}^{\mathrm{MAS}}$ and $\Gamma_{\mathrm{V,M}}^{\mathrm{MAS}}$ of the tangent spaces of the Grassmann manifolds, followed by exponential mappings ($\exp_{\overline{U}}$ and $\exp_{\overline{V}}$, about the Karcher means) to obtain $\mathbf{U}_{\mathrm{M}}^{\mathrm{MAS}}$ and $\mathbf{V}_{\mathrm{M}}^{\mathrm{MAS}}$, and reverse SVD to reconstruct the full strain field.

## 3.3. Surrogate modeling

Having obtained a set of 300 realizations of the input random vector i.e., $\mathcal{X}_{\mathrm{M}} = \left\{ \boldsymbol{x}_{\mathrm{M}}^{(1)}, \boldsymbol{x}_{\mathrm{M}}^{(2)}, \ldots, \boldsymbol{x}_{\mathrm{M}}^{(300)} \right\}$, the 2D subject-specific computational head model is run to compute the corresponding model output solutions, $\mathcal{Y}_{\mathrm{M}}^{\mathrm{MAS}} = \left\{ \boldsymbol{y}_{\mathrm{M,1}}^{\mathrm{MAS}}, \boldsymbol{y}_{\mathrm{M,2}}^{\mathrm{MAS}}, \ldots, \boldsymbol{y}_{\mathrm{M,300}}^{\mathrm{MAS}} \right\}$ and $\mathcal{Y}_{\mathrm{M}}^{\mathrm{CMPS}} = \left\{ \boldsymbol{y}_{\mathrm{M,1}}^{\mathrm{CMPS}}, \boldsymbol{y}_{\mathrm{M,2}}^{\mathrm{CMPS}}, \ldots, \boldsymbol{y}_{\mathrm{M,300}}^{\mathrm{CMPS}} \right\}$, where $\boldsymbol{y}_{\mathrm{M,}i}^{\mathrm{MAS}}$ and $\boldsymbol{y}_{\mathrm{M,}i}^{\mathrm{CMPS}}$, $i = 1, \ldots, 300$, are vectors containing 2125 MAS and 6372 CMPS values, respectively. The goal now is to train two surrogate models using this available input-output data that will map the high-dimensional ($\mathbb{R}^{16}$) input to the very high-dimensional output space of MAS ($\mathbb{R}^{2125}$) and CMPS ($\mathbb{R}^{6372}$) fields. Two major challenges for the surrogate model development are the limited number of training data (due to high cost of the computational head model) and the very high-dimensionality of the outputs, both of which lead to poor accuracy of surrogate models in performing out-of-sample predictions (which is required for UQ). To overcome these challenges, a recently proposed manifold learning-based surrogate framework [48] is employed in this work for developing data-driven surrogate models on very high-dimensional outputs of the computational head models.

The following subsections discuss the surrogate model construction, which is performed in three steps (see Fig. 4): (i) the very high-dimensional solution space is embedded onto low-dimensional Grassmannian diffusion manifold, (ii) a Gaussian process surrogate is trained to map the input space to the low-dimensional solution space, and (iii) the predicted low-dimensional reduced solutions are expanded via geometric harmonics models to reconstruct full, very high-dimensional solutions. It is important to note that in the original framework [48], a Polynomial Chaos Expansion (PCE) was used to create the mapping from the input to the reduced space. However, since PCE suffers from the curse of dimensionality (i.e., exponential growth of the number of unknown coefficients with the input dimension), and in the 2D head model, the input space is high dimensional (16 dimensions), Gaussian process regression is selected as the tool to perform the mapping [27]. Note that several other machine learning methods can also be used for this purpose, such as geometric harmonics [28] and artificial neural networks [49]. For clarity, only the MAS strain field output is considered in the following description; the same procedure is followed for building the surrogate model for mapping input to the CMPS strain field.

### 3.3.1. Step I

In the first step, the very high-dimensional output data is projected onto a low-dimensional space using Grassmannian diffusion maps [26]. For this, each output is first reshaped into matrix form (i.e., $\{ \boldsymbol{y}_{\mathrm{M},i}^{\mathrm{MAS}} \in \mathbb{R}^{2125} \}_{i=1}^{300} \rightarrow \{ \mathbf{F}_{\mathrm{M},i}^{\mathrm{MAS}} \in \mathbb{R}^{85 \times 25} \}_{i=1}^{300}$), and then, linearly projected onto a Grassmann manifold by means of thin singular value decompositions (SVD):

$$\mathbf{F}_{\mathrm{M},i}^{\mathrm{MAS}} = \mathbf{U}_{\mathrm{M},i}^{\mathrm{MAS}} \, \boldsymbol{\Sigma}_{\mathrm{M},i}^{\mathrm{MAS}} \, \mathbf{V}_{\mathrm{M},i}^{\mathrm{MAS}^{\mathrm{T}}} \tag{5}$$

where $\mathbf{U}_{\mathrm{M},i}^{\mathrm{MAS}} \in \mathbb{R}^{85 \times p_i}$, $\boldsymbol{\Sigma}_{\mathrm{M},i}^{\mathrm{MAS}} \in \mathbb{R}^{p_i \times p_i}$, and $\mathbf{V}_{\mathrm{M},i}^{\mathrm{MAS}} \in \mathbb{R}^{25 \times p_i}$; $p_i$ is the rank of $\mathbf{F}_{\mathrm{M},i}^{\mathrm{MAS}}$. For recasting the solutions into matrix form, a general rule-of-thumb is to make the matrix as close to square as possible [50]. Equation (5) factorizes $\mathbf{F}_{\mathrm{M},i}^{\mathrm{MAS}}$ in two orthonormal matrices, $\mathbf{U}_{\mathrm{M},i}^{\mathrm{MAS}}$ and $\mathbf{V}_{\mathrm{M},i}^{\mathrm{MAS}}$, which represent points on two distinct Grassmann manifolds:

$$\mathcal{G}_U \equiv \mathcal{G}(p, 85) = \left\{ \mathrm{span}\left( \mathbf{U}_{\mathrm{M}}^{\mathrm{MAS}} \right) : \mathbf{U}_{\mathrm{M}}^{\mathrm{MAS}} \in \mathbb{R}^{85 \times p} \right\} \tag{6a}$$

$$\mathcal{G}_V \equiv \mathcal{G}(p, 25) = \left\{ \mathrm{span}\left( \mathbf{V}_{\mathrm{M}}^{\mathrm{MAS}} \right) : \mathbf{V}_{\mathrm{M}}^{\mathrm{MAS}} \in \mathbb{R}^{25 \times p} \right\} \tag{6b}$$

where $p = max(p_i)$, $i = 1, \ldots, 300$. Matrix $\boldsymbol{\Sigma}_{\mathrm{M},i}^{\mathrm{MAS}} \in \mathbb{R}^{p \times p}$ is diagonal containing the most important singular values.

Next, diffusion maps are applied to reveal the latent structure of the data on the Grassmann manifolds. Note that unlike commonly employed diffusion maps in a Euclidean space [8,51], a Grassmannian kernel [50] is employed in this case to define similarity between points on the Grassmann manifold. The overall kernel matrix $K\left( \mathbf{U}_{\mathrm{M}}^{\mathrm{MAS}}, \mathbf{V}_{\mathrm{M}}^{\mathrm{MAS}} \right)$ is defined as

$$K\left( \mathbf{U}_{\mathrm{M}}^{\mathrm{MAS}}, \mathbf{V}_{\mathrm{M}}^{\mathrm{MAS}} \right) = K\left( \mathbf{U}_{\mathrm{M}}^{\mathrm{MAS}} \right) \circ K\left( \mathbf{V}_{\mathrm{M}}^{\mathrm{MAS}} \right) \tag{7}$$

where

$$K\left(\mathbf{U}_{\mathrm{M}}^{\mathrm{MAS}}\right): \mathcal{G}_U \times \mathcal{G}_U \rightarrow \mathbb{R}^{300 \times 300} \tag{8a}$$

$$K\left(\mathbf{V}_{\mathrm{M}}^{\mathrm{MAS}}\right): \mathcal{G}_V \times \mathcal{G}_V \rightarrow \mathbb{R}^{300 \times 300} \tag{8b}$$

are individual projection kernels defined on manifolds $\mathcal{G}(p, 85)$ and $\mathcal{G}(p, 25)$, respectively, and $\circ$ denotes the Hadamard product. Finally, for each data point $i$, diffusion coordinates $\boldsymbol{\Theta}_{\mathrm{M},i}^{\mathrm{MAS}} \in \mathbb{R}^q$ are obtained, $q$ being the number of eigenvectors required to reveal the intrinsic structure of the data on the Grassmann manifolds. In this work, $q = 3$ was found to be sufficient to capture the essential features and geometric structure of both the MAS and CMPS fields.

### 3.3.2. Step II

In this step, two Gaussian processes are constructed:

$$\mathcal{GP}_{\boldsymbol{\Theta}}^{\mathrm{MAS}}: \boldsymbol{X}_{\mathrm{M}} \rightarrow \boldsymbol{\Theta}_{\mathrm{M}}^{\mathrm{MAS}} \tag{9a}$$

$$\mathcal{GP}_{\Sigma}^{\mathrm{MAS}}: \boldsymbol{X}_{\mathrm{M}} \rightarrow \boldsymbol{\Sigma}_{\mathrm{M}}^{\mathrm{MAS}} \tag{9b}$$

where $\mathcal{GP}_{\boldsymbol{\Theta}}^{\mathrm{MAS}}$ and $\mathcal{GP}_{\Sigma}^{\mathrm{MAS}}$ map the input random vector $\boldsymbol{X}_{\mathrm{M}}$ to the space of diffusion coordinates $\boldsymbol{\Theta}_{\mathrm{M}}^{\mathrm{MAS}}$ and singular values $\boldsymbol{\Sigma}_{\mathrm{M}}^{\mathrm{MAS}}$, respectively. The mathematical formulations of these Gaussian processes are

$$\mathcal{GP}_{\boldsymbol{\Theta}}^{\mathrm{MAS}}(\boldsymbol{X}_{\mathrm{M}}) = \mathcal{F}_{\boldsymbol{\Theta}}(\boldsymbol{X}_{\mathrm{M}}) + z_{\boldsymbol{\Theta}}(\boldsymbol{X}_{\mathrm{M}}, \omega) \tag{10a}$$

$$\mathcal{GP}_{\Sigma}^{\mathrm{MAS}}(\boldsymbol{X}_{\mathrm{M}}) = \mathcal{F}_{\Sigma}(\boldsymbol{X}_{\mathrm{M}}) + z_{\Sigma}(\boldsymbol{X}_{\mathrm{M}}, \omega) \tag{10b}$$

where $\mathcal{F}_{\boldsymbol{\Theta}, \Sigma}(\boldsymbol{X}_{\mathrm{M}})$ represent the mean value (assumed to be constant in this study) and $z_{\boldsymbol{\Theta}, \Sigma}(\boldsymbol{X}_{\mathrm{M}}, \omega)$ represent zero-mean, stationary Gaussian processes, each defined in terms of a correlation function. Training of a Gaussian process involves calibrating the hyperparameters of this correlation function. In this work, a Gaussian correlation function is selected. Finally, for a new realization $\boldsymbol{x}_{\mathrm{M}}^*$ of the input parameter vector, each trained Gaussian process will return:

$$\mathcal{GP}_{\boldsymbol{\Theta}}^{\mathrm{MAS}}(\boldsymbol{x}_{\mathrm{M}}^*) = \widetilde{\boldsymbol{\Theta}}_{\mathrm{M}}^{\mathrm{MAS}} \tag{11a}$$

$$\mathcal{GP}_{\Sigma}^{\mathrm{MAS}}(\boldsymbol{x}_{\mathrm{M}}^*) = \widetilde{\boldsymbol{\Sigma}}_{\mathrm{M}}^{\mathrm{MAS}} \tag{11b}$$

### 3.3.3. Step III

Training of the Gaussian processes enables the prediction of the low-dimensional diffusion coordinates and singular values for any new realization of the input random vector. However, as the present study is interested in the behavior of the $\mathbb{R}^{2125}$- and $\mathbb{R}^{6372}$-valued MAS and CMPS strain fields, the predicted reduced-order solutions must be mapped back to the original, very high dimensional space: this work employs a method introduced in [52] based on geometric harmonics [28] for this purpose.

First, the mappings between the low-dimensional diffusion manifold and the Grassmann manifolds $\mathcal{G}_U(p, 85)$ and $\mathcal{G}_V(p, 25)$ are defined. For this, $n_c$ clusters of the diffusion coordinates obtained from Step 1 (i.e., $\{\boldsymbol{\Theta}_{\mathrm{M},1}^{\mathrm{MAS}}, \boldsymbol{\Theta}_{\mathrm{M},2}^{\mathrm{MAS}}, \dots, \boldsymbol{\Theta}_{\mathrm{M},300}^{\mathrm{MAS}}\}$) are identified using the $k$-means algorithm. The optimum number of clusters is identified using an adaptive clustering method defined in [52]. Next, for the diffusion coordinates $\widetilde{\boldsymbol{\Theta}}_{\mathrm{M}}^{\mathrm{MAS}}$, the nearest cluster $\{\boldsymbol{\Theta}_{\mathrm{M},j}^{\mathrm{MAS}}\}$ is identified ($j$ contains indices of the diffusion coordinates of the identified cluster). The projections of the corresponding Grassmann points $\{\mathbf{U}_{\mathrm{M},j}^{\mathrm{MAS}}\}$ and $\{\mathbf{V}_{\mathrm{M},j}^{\mathrm{MAS}}\}$ on the tangent spaces with origins at the Karcher means are $\{\boldsymbol{\Gamma}_{\mathrm{U},\mathrm{M},j}^{\mathrm{MAS}}\}$ and $\{\boldsymbol{\Gamma}_{\mathrm{V},\mathrm{M},j}^{\mathrm{MAS}}\}$, respectively (this type of projection from Grassmann manifold to a tangent space is referred to as logarithmic mapping [20]). Using this data, two geometric harmonics models are constructed:

$$f_{\mathbf{U}}^{\mathrm{MAS}} \colon \boldsymbol{\Theta}_{\mathrm{M}}^{\mathrm{MAS}} \in \mathbb{R}^q \to \boldsymbol{\Gamma}_{\mathrm{U},\mathrm{M}}^{\mathrm{MAS}} \in \mathbb{R}^{85 \times p} \tag{12a}$$

$$f_{\mathbf{V}}^{\mathrm{MAS}} \colon \boldsymbol{\Theta}_{\mathrm{M}}^{\mathrm{MAS}} \in \mathbb{R}^q \to \boldsymbol{\Gamma}_{\mathrm{V},\mathrm{M}}^{\mathrm{MAS}} \in \mathbb{R}^{25 \times p} \tag{12b}$$

where $f_{\mathbf{U}}^{\mathrm{MAS}}$ and $f_{\mathbf{V}}^{\mathrm{MAS}}$ define mappings between diffusion coordinates and the corresponding matrices $\boldsymbol{\Gamma}_{\mathrm{U},\mathrm{M}}^{\mathrm{MAS}}$ and $\boldsymbol{\Gamma}_{\mathrm{V},\mathrm{M}}^{\mathrm{MAS}}$, respectively. A Gaussian or RBF kernel is selected for the geometric harmonics models. In a similar manner, the corresponding geometric harmonics models for the CMPS strain field output are also constructed. With the predicted tangent space points $\bar{\boldsymbol{\Gamma}}_{\mathrm{U},\mathrm{M}}^{\mathrm{MAS}}$ and $\bar{\boldsymbol{\Gamma}}_{\mathrm{V},\mathrm{M}}^{\mathrm{MAS}}$ (origins at the Karcher means) for a given $\widetilde{\boldsymbol{\Theta}}_{\mathrm{M}}^{\mathrm{MAS}}$, the corresponding points $\widetilde{\mathbf{U}}_{\mathrm{M}}^{\mathrm{MAS}}$ and $\widetilde{\mathbf{V}}_{\mathrm{M}}^{\mathrm{MAS}}$ on Grassmann manifolds can be obtained using exponential mapping [20] (a reverse logarithmic mapping).

The mappings defined in Eqs. (9) (i.e., Gaussian process surrogates) and (12) (i.e., geometric harmonics models) form the overall data-driven surrogate model, i.e., $\widetilde{\mathcal{M}}^{\mathrm{MAS}} \colon \boldsymbol{X}_{\mathrm{M}} \to \boldsymbol{Y}_{\mathrm{M}}^{\mathrm{MAS}}$, which is used in this study to approximate the 2D subject-specific computational head model. For any given realization of the input random vector $\boldsymbol{X}_{\mathrm{M}}$, the two Gaussian process surrogates in Eq. (9) predict the low-dimensional reduced outputs (i.e., diffusion coordinates of the low-dimensional latent space and the SVD diagonal matrix), while the two geometric harmonics models in Eq. (12) predict the points on the tangent spaces corresponding to the diffusion coordinates, which can be mapped (using exponential mapping) to yield the two SVD orthogonal matrices. The three matrices $\mathbf{U}, \boldsymbol{V},$ and $\boldsymbol{\Sigma}$ can then be combined as in Eq. (5) to yield the matrix version of the output, which is then recast as a column vector to yield the full, very high-dimensional vector strain output (corresponding to the random vector of MAS strain field, $\boldsymbol{Y}_{\mathrm{M}}^{\mathrm{MAS}}$). The corresponding data-driven surrogate model for predicting CMPS strain fields, i.e., $\widetilde{\mathcal{M}}^{\mathrm{CMPS}} \colon \boldsymbol{X}_{\mathrm{M}} \to \boldsymbol{Y}_{\mathrm{M}}^{\mathrm{CMPS}}$, also works in a similar way.

All the calculations presented in this work are performed using two open-source Python packages: UQpy [53] for Grassmannian diffusion maps and Gaussian processes, and *datafold* [54] for geometric harmonics.

## 3.4. Surrogate model validation

To assess the predictive ability of the data-driven surrogate models, leave-one-out cross-validation (LOO-CV) is conducted. In this method, 300 (i.e., full training dataset size) $\widetilde{\mathcal{M}}^{\mathrm{MAS}}$ and $\widetilde{\mathcal{M}}^{\mathrm{CMPS}}$ surrogate models are constructed, each trained using 299 input-output pairs from $\mathcal{X}_{\mathrm{M}}$ and $\mathcal{Y}_{\mathrm{M}}^{\mathrm{MAS}}/\mathcal{Y}_{\mathrm{M}}^{\mathrm{CMPS}}$, and applied to predict the output vector corresponding to the remaining realization of the input random vector (that is not used in the training process). The accuracy of the 300 models are evaluated using the scalar metric of coefficient of determination ($R^2$). The mean and standard deviation of this metric are reported. Mathematically, $R^2$ for a model tested on realization $\boldsymbol{x}_{\mathrm{M}}^*$ is given by

$$R^2 = 1 - \frac{\sum_j \left( \widetilde{\mathcal{M}}_j(x_M^*) - \mathcal{M}_j(x_M^*) \right)^2}{\sum_j \left( \widetilde{\mathcal{M}}_j(x_M^*) - \bar{\mathcal{M}}(x_M^*) \right)^2} \tag{13}$$

where $\mathcal{M}_j(x_M^*)$ is the $j^{th}$ component of the vector output (i.e., MAS or CMPS at a particular brain voxel) obtained from the computational simulation of a 2D subject-specific head model with input material properties $x_M^*$, and $\widetilde{\mathcal{M}}_j(x_M^*)$ is the corresponding component of the vector output predicted by the data-driven surrogate model for the same set of material properties. $\bar{\mathcal{M}}(x_M^*)$ denotes the mean value of $\mathcal{M}_j(x_M^*)$ over all $j$. Using Eq. (13), the accuracy of both the surrogate models, $\widehat{\mathcal{M}}^{MAS}$ (for MAS strain field output) and $\widehat{\mathcal{M}}^{CMPS}$ (for CMPS strain field output) can be computed: for the former, $j$ ranges from 1 to 2125 (i.e., number of white matter brain voxels in CC and CR), while for the latter, it ranges from 1 to 6372 (i.e., total number of brain voxels).

In addition to $R^2$, which is used to measure the overall accuracy of the surrogate model, a second scalar metric of absolute relative error $\epsilon$ is employed to evaluate the local error at individual brain voxels (i.e., at a given $j$),

$$\epsilon_j = \left| \frac{\widetilde{\mathcal{M}}_j(x_M^*) - \mathcal{M}_j(x_M^*)}{\mathcal{M}_j(x_M^*)} \right| \tag{14}$$

## 3.5. Uncertainty propagation

Once the Gaussian process surrogates and geometric harmonics models are trained using the experimental design of 300 realizations and the resulting data-driven surrogate model is validated, the remaining 10,000 realizations of the input random vector are used for uncertainty propagation (note, a total of 10,300 realizations were obtained after the data-driven sampling in Section 3.2). This set of realizations is defined as $\mathcal{X}_{M,UP} = \left\{ x_M^{(301)}, x_M^{(302)}, \dots, x_M^{(10300)} \right\}$. Using the trained Gaussian process surrogates, $\mathcal{GP}_\Theta^{MAS}$ and $\mathcal{GP}_\Sigma^{MAS}$, the corresponding low-dimensional reduced outputs, $\left\{ \widetilde{\Theta}_{M,i}^{MAS} \right\}_{i=1}^{10000}$ (where $\widetilde{\Theta}_{M,i}^{MAS} \in \mathbb{R}^q$) and $\left\{ \widetilde{\Sigma}_{M,i}^{MAS} \right\}_{i=1}^{10000}$ (where $\widetilde{\Sigma}_{M,i}^{MAS} \in \mathbb{R}^{p \times p}$), are predicted. Next, using the trained geometric harmonics models, $f_U^{MAS}$ and $f_V^{MAS}$, the set of matrices $\left\{ \widetilde{\Gamma}_{U,M,i}^{MAS} \right\}_{i=1}^{10000}$ and $\left\{ \widetilde{\Gamma}_{V,M,i}^{MAS} \right\}_{i=1}^{10000}$ are predicted, which yield matrices $\left\{ \widetilde{U}_{M,i}^{MAS} \right\}_{i=1}^{10000}$ and $\left\{ \widetilde{V}_{M,i}^{MAS} \right\}_{i=1}^{10000}$ via exponential mapping. The matrix versions of the final MAS strain output, $\left\{ \widetilde{F}_{M,i}^{MAS} \right\}_{i=1}^{10000}$, are then recovered using Eq. (5). Finally, the matrix solutions are recast to column vectors, i.e., $\left\{ \widetilde{F}_{M,i}^{MAS} \in \mathbb{R}^{85 \times 25} \right\}_{i=1}^{10000} \rightarrow \left\{ \widetilde{y}_{M,i}^{MAS} \in \mathbb{R}^{2125} \right\}_{i=1}^{10000}$, yielding the full, very high-dimensional solutions for all the 10,000 realizations of the input random vector of visco-hyperelastic material properties. Different statistical measures of $Y_M^{MAS}$ are computed from these 10,000 output vectors (e.g., mean and standard deviation). In a similar manner, UQ is conducted for the CMPS strain field via 10,000 evaluations of the $\widehat{\mathcal{M}}^{CMPS}$ model (composed of $\mathcal{GP}_\Theta^{CMPS}$, $\mathcal{GP}_\Sigma^{CMPS}$, $f_U^{CMPS}$ and $f_V^{CMPS}$).

## 4. Results

## 4.1. Model input realizations

As outlined in Section 3.2, a manifold learning-based data-driven method is employed in two steps to generate realizations of the input random vector for the 2D subject-specific head model. In the first step, 900 realizations of the input random vectors corresponding to each of the four brain substructures (i.e., DG, CG, CC and CR) are individually generated. Figure 5(a) shows pair-wise correlation plots of the material properties in such realizations along with those in the available data for a representative brain substructure: the corona radiata (CR). The corona radiata is one of the largest substructures in the brain by volume (and therefore by the number of voxels). Since there are four material properties associated with each substructure, six scatter plots (i.e., $C(4,2)$ combinations) are presented. The red data-points represent the 2534 available material property data points for this region after the removal of outliers (criteria: three scaled median absolute deviations), and the blue data-points represent the generated 900 realizations. Comparisons of the normalized histograms of the individual material parameters are also shown in the figure (within the green boxes). From these plots, it is clear that the new realizations (after the first step) follow the distribution of the original data, and thus it can be said that they are concentrated in subset $\mathcal{S}_4^{CR} \subset \mathbb{R}^4$ on which the probability distribution of the material properties of this substructure is concentrated. Similar agreement with the distribution of available data is seen for the other three brain substructures as well (not shown). With the realizations of all the four substructures, in the second step, 300 realizations of the sixteen-dimensional input random vector of material properties for the entire head model are generated for training the data-driven surrogate model. Representation of all these input random vector components via pair-wise correlation plots would require 120 scatter plots (i.e., $C(16,2)$). For brevity, Figure 5(b) shows pair-wise correlation plots and normalized histograms of four material properties: one selected from each substructure. Again, it is seen that the 300 new realizations (in blue) follow the distribution of the original 900 realizations from individual brain substructures (in red) that were generated in the first step (note that 100 out of these 900 realizations were used as inputs for the data-driven generator of realizations in the second step). Thus, the generated realizations for the full head model are concentrated in the subset (initially unknown) $\mathcal{S}_{16} \subset \mathbb{R}^{16}$ on which the probability distribution of the sixteen input random vector components is concentrated.

Finally, note that the space-filling ability of this data-driven method is proportional to the number of generated realizations. For example, normalized histograms (as in Fig. 5(b)) for a hypothetical case when only 50 new realizations are generated for training the surrogate model (instead of 300) show a much greater mismatch between the distributions of the new realizations and the original 900 realizations generated from individual brain substructures (see Fig. S2 in the supplementary material).

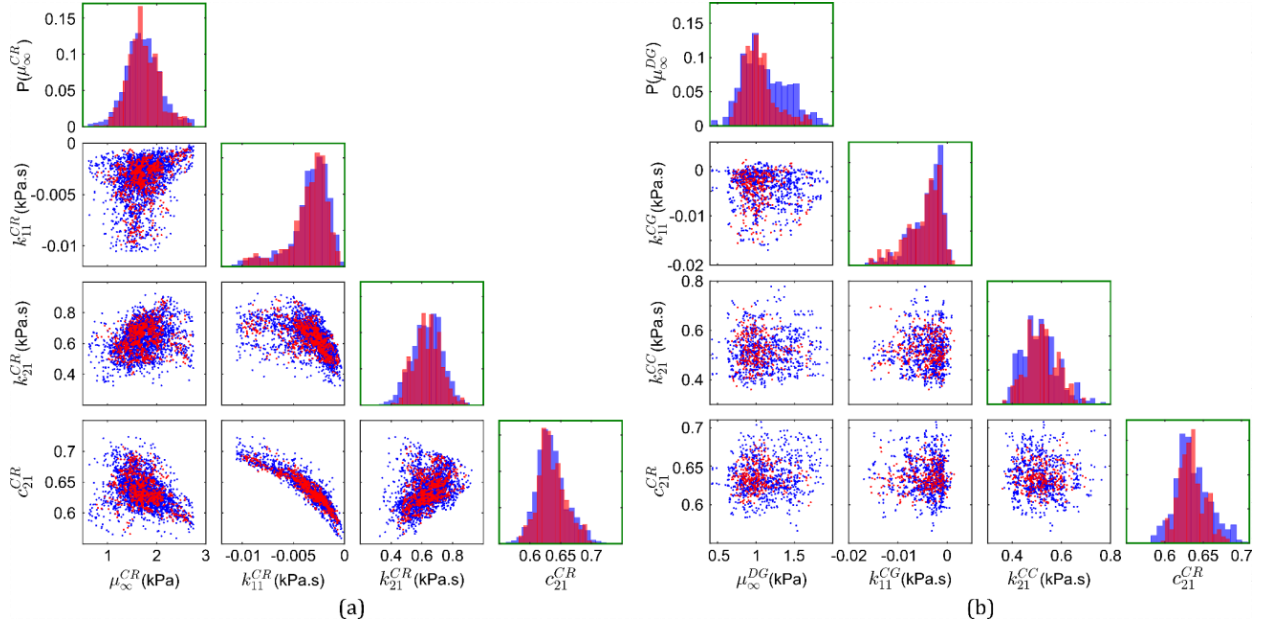

**Figure 5**. Pairwise correlation plots and normalized histograms of (a) four material properties of the corona radiata substructure: available data from MRE (in blue) and 900 generated realizations for this substructure (in red), and (b) four representative material properties from the four brain substructures: 900 realizations from individual substructures (in blue) and 300 new realizations of the overall 16D input random vector for the head model (in red).

## 4.2. Surrogate model performance

The 2D subject-specific head model simulations of the 300 input realizations provide the input-output (i.e., MAS or CMPS strain) pairs used as training-testing data for the data-driven surrogate models. Figure 6(a) compares the MAS strain field predicted by a representative surrogate model trained using 299 input-output pairs with the one obtained from head model simulation, for a particular set of material properties not part of the training input dataset. Note that these 2D fields are created by assigning each component of the strain vector output the spatial location (in the x-y plane) of a particular brain voxel; the correspondence between brain voxel location and strain vector components is a priori fixed. From this figure, a good agreement is observed between the computational model and the surrogate model for MAS strain. Figure 6(b) shows the spatial distribution of the absolute relative error metric $\epsilon$ (Eq. (14)), revealing a very high accuracy ($\epsilon \leq 0.1$) in regions of relatively high MAS, but a low accuracy ($\epsilon \geq 1$) in regions associated with negligible MAS. The latter is expected because in regions of negligible MAS, even a small discrepancy on the order of 0.001 mm/mm strain between the two models can lead to a very high absolute "relative" error. Overall, this representative surrogate model results in an excellent $R^2$-value of 0.95 (calculated over all the MAS voxels). Figure 6(c) shows the corresponding comparison of CMPS strain fields generated by the computational model and the representative surrogate model of CMPS strain, with the spatial distribution of $\epsilon$ shown in Fig. 6(d). Again, a good agreement is observed, this time, with a relatively uniform distribution of $\epsilon$. For this representative surrogate model, an $R^2$-value of 0.91 is obtained over all the brain voxels.

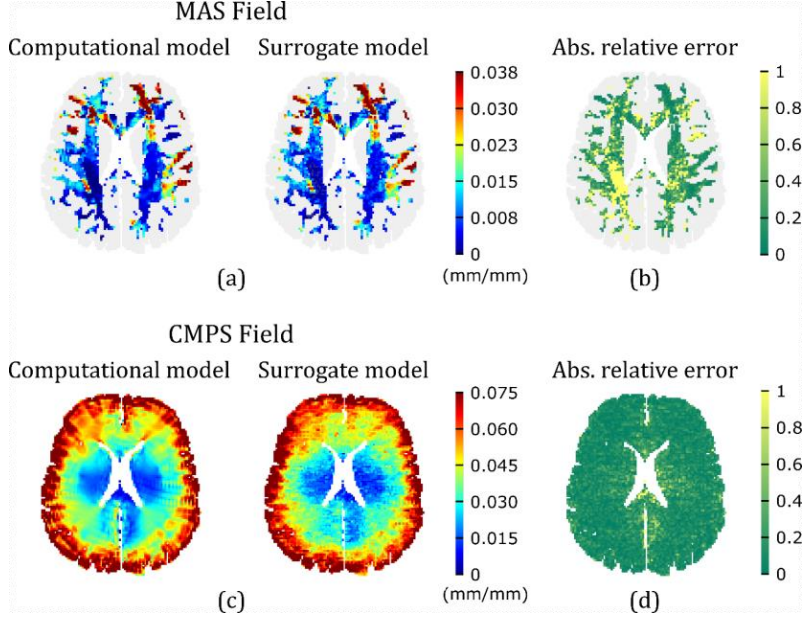

**Figure 6**. Comparison of predicted strain fields for a representative realization of the input random vector (i.e., $\boldsymbol{x}_M^o$ = (0.744, -0.158, 0.467, 0.908, 0.915, -0.450×10⁻², 0.239, 0.716, 2.05, -0.141×10⁻², 0.643, 0.618, 1.66, -0.181×10⁻², 0.513, 0.619); see order of material properties in Eq. (3), and units in Table 1) by surrogate models trained using the remaining 299 input-output pairs, with the corresponding fields resulted from the computational head model: (a) MAS strain field, (c) CMPS strain field. The corresponding absolute relative errors for the MAS and CMPS strain fields are shown in (b) and (d), respectively.

Overall, for the two data-driven surrogate models developed in this work for MAS and CMPS strain fields, LOO-CV results in an $R^2$ of 0.93 ± 0.03 (mean ± standard deviation) for the MAS case, and 0.90 ± 0.02 for the CMPS case. In addition to reasonably approximating the computational head model with a small training dataset, the surrogate models also lead to a significant reduction in computational cost. For the representative models in Fig. 6, the two surrogate model runs for predicting MAS and CMPS strain fields completed in 0.010 ± 0.002 seconds (based on 5 runs) on a personal computer, while the corresponding MPM simulation of the computational model completed in 9948.64 seconds on an HPC cluster (with 72 parallel tasks). Thus, the surrogate model offers more than a million times faster performance.

In general, the overall accuracy of a surrogate model (e.g., $R^2$) improves with the training dataset size, eventually reaching an asymptote for a sufficiently large training dataset [55]. In this study, it was seen that the mean $R^2$ values for both surrogate models reached asymptotic values at an approximate training data size of 100 input-output pairs (even with only 25 input-output pairs, mean $R^2$ values of greater than 0.80 were obtained). However, as noted in the previous subsection, input parameters from such small training dataset are in a relatively poor agreement with the distribution of the original in-vivo MRE data (i.e., poor space-filling): this was the reason behind choosing the training dataset size of 300 for surrogate model development in this work.

### 4.3. Uncertainty in the strain-response of the 2D subject-specific model

The trained surrogate models are used to predict strain outputs corresponding to the 10,000 additional input realizations (see Section 3.5) for uncertainty quantification of the computational head model. In this regard, uncertainty in the $p^{\text{th}}$-percentile strain is first analyzed ($p^{\text{th}}$-percentile strain of a given predicted strain field is the smallest strain value that is greater than or equal to $p$ percent of all the strain values in that field). Figure 7(a) plots the $p^{\text{th}}$-percentile MAS (denoted by MAS$p$) for the 2D subject-specific head model investigated in this work in the $p \in [5,100]$ range: data-points represent the average values, while error bars represent one standard deviation. The corresponding $p^{\text{th}}$-percentile CMPS (denoted by CMPS$p$) are shown in Fig. 7(b). From these figures, the absolute values of standard deviations for both MAS$p$ and CMPS$p$ increase with the percentile score. For a better insight into the relative uncertainty of strain outputs, Fig. 7(c) plots the coefficient of variation (CV), defined as the ratio of the standard deviation to the mean, for the two strain outputs as a function of the percentile score. For MAS$p$ values evaluated at less than 20-percentile, CV is greater than 1/3, i.e., the mean is less than three times the standard deviation. This is equivalent to a signal-to-noise ratio [56] (i.e., the reciprocal of CV) of greater than 3. Thus, for low percentiles, the MAS$p$ strain output is associated with a high uncertainty; this is expected because of the very small mean strain values (denominator in the CV formula) at low percentile scores, even if the absolute standard deviation values remain reasonable. With increasing percentile score, the CV of MAS$p$ decreases and remains below the 1/3 level. Unlike MAS$p$, CMPS$p$ (which have relatively larger mean values), the CV at all the investigated percentile scores is less than 1/3, suggesting a consistently low uncertainty. Similar to MAS$p$, the CV in the case of CMPS$p$ tends to become very large as $p$ tends to 0 (because then, the predicted mean strain values are very small).

Note that for both MAS$p$ and CMPS$p$, a low CV is observed in the 50- to 95-percentile score range. As 50- and 95-percentile MAS and CMPS are commonly employed in the brain biomechanics community for the prediction and vulnerability assessment to brain injury [6,33,57], the uncertainty in these brain injury metrics (also called brain injury predictor variables) for subject-specific models is reasonably low (actual values of mean, standard deviation and CV are listed in Table 2). Due to their insignificant effect on brain injury prediction, strains evaluated at percentile scores below 50 hold lesser importance.

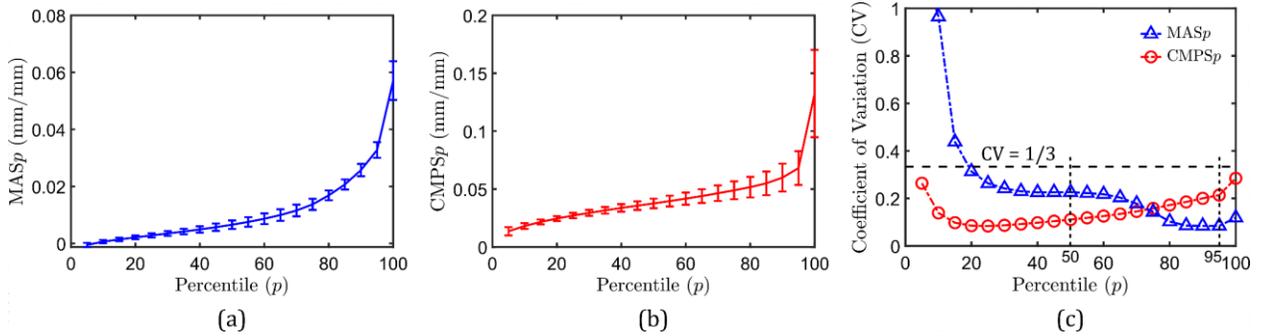

(a)  (b)  (c)

**Figure 7**. Uncertainty in the $p^{\text{th}}$ percentile MAS (MAS$p$) and $p^{\text{th}}$ percentile CMPS (CMPS$p$) outputs of the 2D head model: (a) MAS$p$ versus $p$, (b) CMPS$p$ versus $p$, and (c) CV of MAS$p$ and CMPS$p$ versus $p$.

To visualize uncertainty in the full-field strain outputs, Figs. 8(a) and 8(b) plot CV at all the voxels in the MAS and CMPS strain fields, respectively, that are associated with average strain greater than the average 50-percentile value. It is seen that the uncertainty in MAS is relatively high (CV > 0.4) in the right brain-hemisphere (note, the right side of the brain is on the left in the figure), and is reasonably low in the left brain-hemisphere (especially in the corona radiata). In the case of the CMPS field, a low uncertainty is observed in the majority of the brain layer area, with relatively high CV near the interface of the cortical gray matter with the SAS. Overall, the CV of the MAS field is $0.29 \pm 0.19$, while that of the CMPS field is $0.20 \pm 0.05$. Thus, compared to the CMPS strain field output, the MAS strain field output is associated with a larger (on average) and highly spatially varying uncertainty.

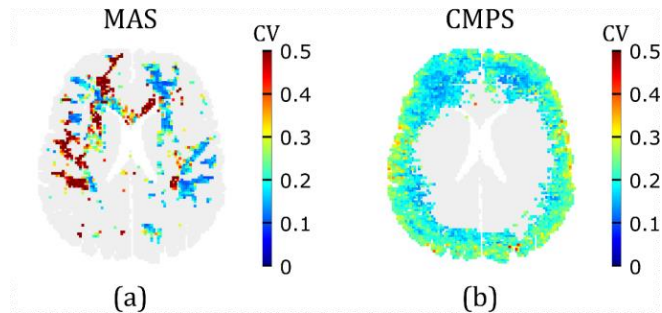

**Figure 8**. Uncertainty in the strain field outputs of the 2D head model: (a) CV in the MAS in brain voxels where MAS is greater than the average 50-percentile MAS (i.e., $\overline{MAS50}$); (b) CV in the CMPS in brain voxels where CMPS is greater than the average 50-percentile CMPS (i.e., $\overline{CMPS50}$).

**Table 2**. Mean and standard deviation of the scalar strain outputs of the 2D subject-specific head model, which are commonly employed as percentile-based brain injury predictor variables.

| Scalar MAS output | Mean | Std. dev. | CV | Scalar CMPS output | Mean | Std. dev. | CV |
|---|---|---|---|---|---|---|---|
| MAS50 (mm/mm) | 0.0066 | 0.0015 | 0.23 | CMPS50 (mm/mm) | 0.0376 | 0.0042 | 0.11 |
| MAS95 (mm/mm) | 0.0328 | 0.0028 | 0.08 | CMPS95 (mm/mm) | 0.0681 | 0.0146 | 0.21 |

Now, the uncertainty in the area fraction of the brain that is associated with strain greater than a certain threshold is analyzed (such area/volume (for 3D models) fractions are sometimes used in metrics for assessing likelihood of injury [6,33]). In this regard, AF-$\overline{MASp}$ denotes the area fraction with MAS greater than the mean $p^{\text{th}}$-percentile MAS threshold (i.e., $\overline{MASp}$), and AF-$\overline{CMPSp}$ denotes the area fraction with CMPS greater than the mean $p^{\text{th}}$-percentile CMPS threshold (i.e., $\overline{CMPSp}$). Figures 9(a) and 9(b) plot AF-$\overline{MASp}$ and AF-$\overline{CMPSp}$ versus $p$, respectively. As expected, smaller (average) area fractions of the brain layer are associated with higher percentile scores (that correspond to larger strain thresholds), leading to monotonically decreasing responses. In the case of AF-$\overline{MASp}$, the absolute value of the standard deviation generally decreases with the percentile score. The CV (see Fig. 9(c)), on the other hand, increases with the percentile score especially in the low and high percentile regimes; nevertheless, it consistently stays below 1/3. Unlike AF-$\overline{MASp}$, AF-$\overline{CMPSp}$ is associated with a much higher uncertainty at percentile thresholds of $p > 65$. For both AF-$\overline{MASp}$ and AF-$\overline{CMPSp}$, the CV tends to infinity as $p$ tends to 100: this is expected because at as $p$ tends to 100, the evaluated mean area fractions (denominator in the CV formula) tend to 0. Overall, among the four investigated scalar strain output measures — MAS$p$, CMPS$p$, AF-$\overline{MASp}$, AF-$\overline{CMPSp}$ — AF-$\overline{CMPSp}$ is associated with the maximum uncertainty.

In real-life head injury models, the area/volume fractions of the brain with MAS and CMPS above certain "fixed" strain thresholds are used as injury predictor variables [6,33]. The values of these strain thresholds vary widely in the literature [6,32]. Since the head model considered in this study was used to simulate non-injurious loading, the available injury thresholds in the literature are generally much higher than the average strain values in this study [5,37,57]. Nevertheless, from Fig. 9, it appears that for general subject-specific head models, the uncertainty in the area fraction-based injury predictor variables will depend on both the magnitude of the strains predicted by the model and the chosen strain threshold value. If a given head model results in very large strains (say, due to high loading severity) such that the chosen strain threshold corresponds to a low percentile value, the area fraction-based injury thresholds will have low uncertainty. However, if the head model under investigation results in relatively small strains such that the chosen strain threshold corresponds to a very high percentile, then the uncertainty associated with the area fraction-based injury thresholds will be high.

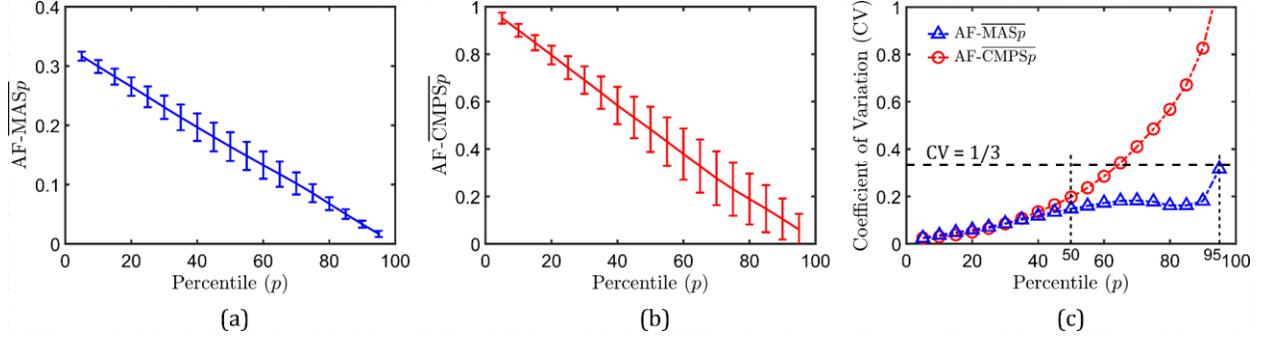

**Figure 9**. Uncertainty in the predicted area fractions (AF-$\overline{MASp}$ and AF-$\overline{CMPSp}$) associated with MAS and CMPS values greater than their mean $p^{\text{th}}$ percentile thresholds: (a) AF-$\overline{MASp}$ versus $p$, (b) AF-$\overline{CMPSp}$ versus $p$, and (c) CV of AF-$\overline{MASp}$ and AF-$\overline{CMPSp}$ versus $p$.

Finally, the spatial uncertainty in the predicted regions where strain exceeds the percentile-based injury predictor variables (i.e., mean values of MAS50, MAS95, CMPS50, and CMPS95, listed in Table 2) is studied. In this regard, four probability values are computed for every brain voxel as

$$P_j^{\text{MAS}>\overline{\text{MAS50}}} = \frac{\text{card}\{\widetilde{\boldsymbol{y}}_{\text{M},i}^{\text{MAS}}(j) > \overline{\text{MAS50}}\}}{10000}, \qquad P_j^{\text{MAS}>\overline{\text{MAS95}}} = \frac{\text{card}\{\widetilde{\boldsymbol{y}}_{\text{M},i}^{\text{MAS}}(j) > \overline{\text{MAS95}}\}}{10000} \qquad (16a)$$

$$P_j^{\text{CMPS}>\overline{\text{CMPS50}}} = \frac{\text{card}\{\widetilde{\boldsymbol{y}}_{\text{M},i}^{\text{CMPS}}(j) > \overline{\text{CMPS50}}\}}{10000}, \qquad P_j^{\text{CMPS}>\overline{\text{CMPS95}}} = \frac{\text{card}\{\widetilde{\boldsymbol{y}}_{\text{M},i}^{\text{CMPS}}(j) > \overline{\text{CMPS95}}\}}{10000} \quad (16b)$$

where $P_j^{\text{MAS}>\overline{\text{MAS}p}}$ ($p = 50, 95$) denotes the fraction of the total number of realizations used for UQ (i.e., 10,000) that predicted MAS value ($\widetilde{\boldsymbol{y}}_{\text{M}}^{\text{MAS}}(j)$) of greater than $\overline{\text{MAS}p}$, at the $j^{th}$ brain voxel. Note, the card$\{\cdot\}$ operator denotes the cardinality (i.e., number of components) of a set; $\widetilde{\boldsymbol{y}}_{\text{M},i}^{\text{MAS}}(j)$ is the MAS value resulted from the $i^{th}$ realization of the surrogate model ($i$ ranges from 1 to 10,000) at the $j^{th}$ brain voxel. In other words, $P_j^{\text{MAS}>\overline{\text{MAS}p}}$ ($p = 50, 95$) is the probability that the predicted MAS value at $j^{th}$ brain voxel is greater than $\overline{\text{MAS}p}$. Similarly, $P_j^{\text{CMPS}>\overline{\text{CMPS}p}}$ denotes the corresponding probability that CMPS at $j^{th}$ brain voxel (i.e., $\widetilde{\boldsymbol{y}}_{\text{M},i}^{\text{CMPS}}(j)$) for the $i^{th}$ realization is greater than $\overline{\text{CMPS}p}$.

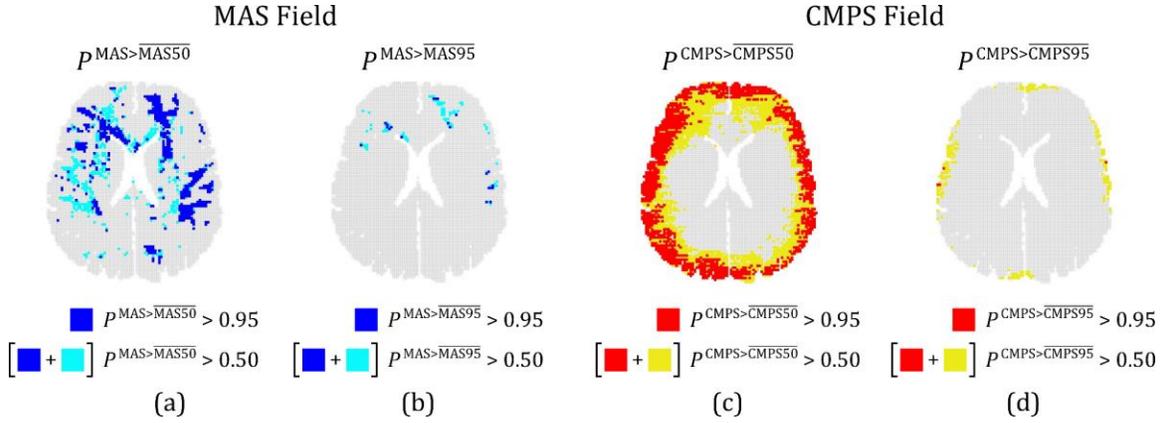

**Figure 10**. (a) Predicted regions where the probability of MAS being greater than $\overline{\text{MAS50}}$ (i.e., $P^{\text{MAS}>\overline{\text{MAS50}}}$), is at least 0.95 (blue) or 0.5 (cyan and blue). Corresponding regions for (b) $P^{\text{MAS}>\overline{\text{MAS95}}}$, (c) $P^{\text{CMPS}>\overline{\text{CMPS50}}}$, and (d) $P^{\text{CMPS}>\overline{\text{CMPS95}}}$. In (c) and (d), regions where the relevant probability measure is at least 0.95 are shown in red, and where it is at least 0.5 is represented by the combined red and yellow regions.

Figure 10(a) shows the predicted area in the brain layer where $P^{\text{MAS}>\overline{\text{MAS50}}}$ is at least 0.5 (combined blue and cyan regions), which corresponds to at least a 50% chance that MAS (at a given brain voxel) is greater than $\overline{\text{MAS50}}$. Comparing this to the considerably smaller region (in blue) with a minimum 95% chance (i.e., $P^{\text{MAS}>\overline{\text{MAS50}}} \geq 0.95$), a considerable uncertainty in the predicted area of the brain with MAS greater than $\overline{\text{MAS50}}$ is evident; this predicted area corresponds to AF-$\overline{\text{MAS50}}$ (see Fig. 9(a)). Similar conclusions can be drawn for the predicted area where MAS is greater than $\overline{\text{MAS95}}$ (see Fig. 10(b)), and for the predicted areas where CMPS is greater than $\overline{\text{CMPS50}}$ and $\overline{\text{CMPS95}}$, respectively (Figs. 10(c-d)). Interestingly, from Fig. 10(d), there are only 4 brain voxels (i.e., 0.06% 2D brain layer area) where at least 95% of the realizations result in a CMPS that is greater than $\overline{\text{CMPS95}}$, even though a 3.40% 2D brain area exists where at least 50% of the realizations result in a CMPS that is greater than $\overline{\text{CMPS95}}$. This significant difference between the predicted regions with CMPS of greater than $\overline{\text{CMPS95}}$ likely resulted in the very high uncertainty (CV > 1) of AF-$\overline{\text{CMPS}p}$ at $p$ = 95 (Fig. 9(c)).

## 5. Summary and Discussion

This work presents a manifold learning-based data-driven framework to quantify the effect of variability and uncertainty in the input parameters of modern biofidelic computational head models on their full-field deformation/strain outputs. This UQ framework is composed of two stages. In the first stage, a data-driven method is used to efficiently sample multiple realizations of the random vector of the input parameter of interest. These realizations are used for training surrogate models in the second stage (low-cost mappings between input and output) and for uncertainty propagation (using the trained surrogate models). The surrogate models employ Grassmannian diffusion maps for dimensionality reduction of the high-dimensional output, and Gaussian process surrogates and geometric harmonics models to create end-to-end mappings between the high-dimensional input and output. This UQ framework is applied to a representative problem of a subject-specific 2D head model of a 31-year-old male, in which the inter-region variability of material properties derived from in-vivo MRE constituted the source of uncertainty, and the outputs of interest were the full MAS and CMPS strain fields. This model is associated with both high-dimensionality of inputs and outputs, and a high computational cost.

From the results, it is seen that the data-driven method for generating realizations resulted in 300 new realizations of the input random vector that are consistent with the distribution of the available material property data from in-vivo MRE. This highlights the ability of this method to accurately discover the a priori unknown probability distribution of the input parameter data. Using the 300 pairs of inputs and outputs generated using the computational model simulations, two data-driven surrogate models (for MAS and CMPS strain output) were trained. The surrogate models approximated the response of the full computational model with very good accuracy; at the same time, these surrogate models provided a huge cost reduction compared to the full computational model. The trained surrogate models allowed efficient uncertainty propagation via 10,000 Monte Carlo simulations that executed in a matter of seconds (for comparison, 10,000 full computational head model simulations would take years). Clearly, the proposed framework overcomes the challenges associated with UQ of computationally expensive, high-dimensional head models.

The UQ of the subject-specific 2D computational head model shows that for both the MAS and CMPS strain outputs, the strain values evaluated at different percentile scores can be associated with very different CV. For very small mean strains at low percentile scores, the uncertainty (i.e., value of CV) tends to be high. On the other hand, in the range of 50- to 95-percentile scores, the uncertainty is reasonably low, which results in a low uncertainty in the percentile-based brain injury predictor variables (i.e., MAS50, MAS95, CMPS50, and CMPS95). In the case of the area fraction of the brain layer with strain greater than a given strain threshold (corresponding to some percentile score), the uncertainty tends to be high for very small area fraction estimates at high percentile scores. On the other hand, for large area fraction estimates at low percentiles scores, the uncertainty is low. Ultimately, it appears that the uncertainty in brain injury predictor variables for a general subject-specific head model depends on the magnitude of the strains produced by the simulations and the chosen value of the strain threshold (for area fraction-based predictors). UQ of computational head models using the proposed framework can guide brain modelers in selecting reliable predictors for assessing the brain injury risk in different loading conditions.

In addition to the scalar strain percentile and area fraction measures, the uncertainty in the full-field strain output of the investigated subject-specific 2D head model is also analyzed. The uncertainty in the MAS strain is shown to vary significantly across the brain layer. On the other hand, the CMPS strain is associated with a relatively homogeneous (spatially) uncertainty. On average, the MAS strain field is associated with a higher uncertainty. Overall, the spatial variation in model uncertainty further highlights the importance of considering full, high-dimensional strain outputs in UQ of head models. In-depth studies on such spatially varying uncertainties can guide specific modifications in the model and improvements in the experiments that provide input data. A considerable uncertainty in the predicted regions where strain is greater than the 50- and 95-percentile strain thresholds is also observed. The uncertainty estimates of the subject-specific 2D computational head model underscore the significance of the influence that the inter-region variability in brain tissue material properties can have on a subject-specific head model's output (both strain fields and the related injury prediction).

The application of the proposed data-driven UQ framework goes far beyond subject-specific head models; this framework can be used to quantify uncertainty for many other input parameter uncertainty and variability cases. For example, for the "average" head models (as opposed to "subject-specific" models that are specific to a particular human subject) that employ head geometry of a 50th-percentile adult male [16], the proposed framework can be used to quantify model uncertainty due to variability in the head geometry (e.g., length, breadth, height, circumference, aspect ratio, volume, etc.) across the human population.

Similarly, the effect of variability in brain tissue material properties across the population can also be quantified. Notably, several recent in-vivo MRE studies have shown significant variability in brain tissue material properties between different ages and genders [58,59]. Finally, as boundary conditions for many head injury models come from measurements that can be associated with high uncertainty (e.g., video analysis of concussive impacts [60]), it will be useful to study the effect of this input parameter (quantified via peak acceleration, loading duration, etc.) on the model output. Ultimately, UQ of computational head models can allow researchers to reliability infer predictions of these models for the better understanding of TBI.

## Acknowledgements

This research was supported by the National Institute of Neurological Disorders and Stroke of the National Institutes of Health (NIH) under Grant No. U01 NS11212. The content is solely the responsibility of the authors and does not necessarily represent the official views of the NIH.

## References

[1]    National Center for Health Statistics: Mortality Data on CDC WONDER, (n.d.). https://wonder.cdc.gov/mcd.html.

[2]    D. Pavlovic, S. Pekic, M. Stojanovic, V. Popovic, Traumatic brain injury: neuropathological, neurocognitive and neurobehavioral sequelae, Pituitary. 22 (2019) 270–282. https://doi.org/10.1007/s11102-019-00957-9.

[3]    N.A. Shaw, The neurophysiology of concussion, Prog. Neurobiol. 67 (2002) 281–344. https://doi.org/10.1016/S0301-0082(02)00018-7.

[4]    D. Sahoo, C. Deck, R. Willinger, Brain injury tolerance limit based on computation of axonal strain, Accid. Anal. Prev. 92 (2016) 53–70. https://doi.org/10.1016/j.aap.2016.03.013.

[5]    E.G. Takhounts, M.J. Craig, K. Moorhouse, J. McFadden, V. Hasija, Development of Brain Injury Criteria (BrIC), in: Stapp Car Crash J., 2013: pp. 243–266. https://doi.org/10.4271/2013-22-0010.

[6]    W. Zhao, Y. Cai, Z. Li, S. Ji, Injury prediction and vulnerability assessment using strain and susceptibility measures of the deep white matter, Biomech. Model. Mechanobiol. 16 (2017) 1709–1727. https://doi.org/10.1007/s10237-017-0915-5.

[7]    P. V. Bayly, A. Alshareef, A.K. Knutsen, K. Upadhyay, R.J. Okamoto, A. Carass, J.A. Butman, D.L. Pham, J.L. Prince, K.T. Ramesh, C.L. Johnson, MR Imaging of Human Brain Mechanics In Vivo: New Measurements to Facilitate the Development of Computational Models of Brain Injury, Ann. Biomed. Eng. 49 (2021) 2677–2692. https://doi.org/10.1007/s10439-021-02820-0.

[8]    R.G. Steen, R.M. Hamer, J.A. Lieberman, Measuring Brain Volume by MR Imaging: Impact of Measurement Precision and Natural Variation on Sample Size Requirements, Am. J. Neuroradiol. 28 (2007) 1119–1125. https://doi.org/10.3174/ajnr.A0537.

[9]    B. Rashid, M. Destrade, M.D. Gilchrist, Mechanical characterization of brain tissue in tension at dynamic strain rates, J. Mech. Behav. Biomed. Mater. 33 (2014) 43–54. https://doi.org/10.1016/j.jmbbm.2012.07.015.

[10] A. TAMURA, S. HAYASHI, K. NAGAYAMA, T. MATSUMOTO, Mechanical Characterization of Brain Tissue in High-Rate Extension, J. Biomech. Sci. Eng. 3 (2008) 263–274. https://doi.org/10.1299/jbse.3.263.

[11] M.T. Prange, S.S. Margulies, Regional, Directional, and Age-Dependent Properties of the Brain Undergoing Large Deformation, J. Biomech. Eng. 124 (2002) 244. https://doi.org/10.1115/1.1449907.

[12] C.L. Johnson, M.D.J. McGarry, A.A. Gharibans, J.B. Weaver, K.D. Paulsen, H. Wang, W.C. Olivero, B.P. Sutton, J.G. Georgiadis, Local mechanical properties of white matter structures in the human brain, Neuroimage. 79 (2013) 145–152. https://doi.org/10.1016/j.neuroimage.2013.04.089.

[13] C.L. Johnson, H. Schwarb, M. D.J. McGarry, A.T. Anderson, G.R. Huesmann, B.P. Sutton, N.J. Cohen, Viscoelasticity of subcortical gray matter structures, Hum. Brain Mapp. 37 (2016) 4221–4233. https://doi.org/10.1002/hbm.23314.

[14] W. Zhao, B. Choate, S. Ji, Material properties of the brain in injury-relevant conditions – Experiments and computational modeling, J. Mech. Behav. Biomed. Mater. 80 (2018) 222–234. https://doi.org/10.1016/j.jmbbm.2018.02.005.

[15] K. Teferra, X.G. Tan, A. Iliopoulos, J. Michopoulos, S. Qidwai, Effect of human head morphological variability on the mechanical response of blast overpressure loading, Int. j. Numer. Method. Biomed. Eng. 34 (2018) e3109. https://doi.org/10.1002/cnm.3109.

[16] D. Sahoo, C. Deck, R. Willinger, Development and validation of an advanced anisotropic visco-hyperelastic human brain FE model, J. Mech. Behav. Biomed. Mater. 33 (2014) 24–42. https://doi.org/10.1016/j.jmbbm.2013.08.022.

[17] S. Ganpule, N.P. Daphalapurkar, K.T. Ramesh, A.K. Knutsen, D.L. Pham, P. V. Bayly, J.L. Prince, A Three-Dimensional Computational Human Head Model That Captures Live Human Brain Dynamics, J. Neurotrauma. 34 (2017) 2154–2166. https://doi.org/10.1089/neu.2016.4744.

[18] S. Ji, W. Zhao, J.C. Ford, J.G. Beckwith, R.P. Bolander, R.M. Greenwald, L.A. Flashman, K.D. Paulsen, T.W. McAllister, Group-Wise Evaluation and Comparison of White Matter Fiber Strain and Maximum Principal Strain in Sports-Related Concussion, J. Neurotrauma. 32 (2015) 441–454. https://doi.org/10.1089/neu.2013.3268.

[19] B. Sudret, S. Marelli, J. Wiart, Surrogate Models for Uncertainty Quantification : An Overview, in: 2017 11th Eur. Conf. Antennas Propag., IEEE, Paris, 2017: pp. 793–797. https://doi.org/10.23919/EuCAP.2017.7928679.

[20] D.G. Giovanis, M.D. Shields, Uncertainty quantification for complex systems with very high dimensional response using Grassmann manifold variations, J. Comput. Phys. 364 (2018) 393–415. https://doi.org/10.1016/j.jcp.2018.03.009.

[21] A.P. Iliopoulos, J.G. Michopoulos, P. Avery, C. Farhat, K. Teferra, S. Qidwai, Towards Model Order Reduction for Uncertainty Propagation in Blast-Induced Traumatic Brain Injury, in: Vol. 1 37th Comput. Inf. Eng. Conf., American Society of Mechanical Engineers, 2017: pp. 1–12.

https://doi.org/10.1115/DETC2017-67556.

[22]  T. Aymard, T. Fogang, Finite Element Simulation of Human Head under Frontal Impact with Uncertainties in Constitutive Modeling and Material Parameters, The University at Buffalo, State University of New York, 2015. https://www.proquest.com/docview/1733327644?pq-origsite=gscholar&fromopenview=true.

[23]  S. Kacker, The Role of Constitutive Model in Traumatic Brain Injury Prediction, University of Cincinnati, 2019. https://etd.ohiolink.edu/apexprod/rws_etd/send_file/send?accession=ucin1563874757653453&disposition=inline.

[24]  M. Hazay, D. Dénes, I. Bojtár, The probability of traumatic brain injuries based on tissue-level reliability analysis., Acta Bioeng. Biomech. 21 (2019) 141–152. https://doi.org/10.5277/ABB-01281-2018-02.

[25]  C. Soize, R. Ghanem, Data-driven probability concentration and sampling on manifold, J. Comput. Phys. 321 (2016) 242–258. https://doi.org/10.1016/j.jcp.2016.05.044.

[26]  K.R.M. dos Santos, D.G. Giovanis, M.D. Shields, Grassmannian diffusion maps based dimension reduction and classification for high-dimensional data, ArXiv. (2020) 1–26. http://arxiv.org/abs/2009.07547.

[27]  C.K.I. Williams, C.E. Rasmussen, Gaussian processes for regression, in: D.S. Touretzky, M.C. Mozer, M.E. Hasselmo (Eds.), Adv. Neural Inf. Process. Syst. 8, MIT, 1996.

[28]  R.R. Coifman, S. Lafon, Geometric harmonics: A novel tool for multiscale out-of-sample extension of empirical functions, Appl. Comput. Harmon. Anal. 21 (2006) 31–52. https://doi.org/10.1016/j.acha.2005.07.005.

[29]  K. Upadhyay, A. Alshareef, A.K. Knutsen, C.L. Johnson, A. Carass, P. V. Bayly, K.T. Ramesh, Development and Validation of Subject-Specific 3D Human Head Models Based on a Nonlinear Visco-Hyperelastic Constitutive Framework, BioRxiv. (2021).

[30]  L. V. Hiscox, C.L. Johnson, E. Barnhill, M.D.J. McGarry, J. Huston, E.J.R. van Beek, J.M. Starr, N. Roberts, Magnetic resonance elastography (MRE) of the human brain: technique, findings and clinical applications, Phys. Med. Biol. 61 (2016) R401–R437. https://doi.org/10.1088/0031-9155/61/24/R401.

[31]  L. V Hiscox, C.L. Johnson, M.D.J. McGarry, M. Perrins, A. Littlejohn, E.J.R. van Beek, N. Roberts, J.M. Starr, High-resolution magnetic resonance elastography reveals differences in subcortical gray matter viscoelasticity between young and healthy older adults, Neurobiol. Aging. 65 (2018) 158–167. https://doi.org/10.1016/j.neurobiolaging.2018.01.010.

[32]  A.K. Knutsen, A.D. Gomez, M. Gangolli, W. Wang, D. Chan, Y. Lu, E. Christoforou, J.L. Prince, P. V Bayly, J.A. Butman, D.L. Pham, In vivo estimates of axonal stretch and 3D brain deformation during mild head impact, Brain Multiphysics. (2020) 100015. https://doi.org/10.1016/j.brain.2020.100015.

[33]  R.W. Carlsen, A.L. Fawzi, Y. Wan, H. Kesari, C. Franck, A quantitative relationship between

rotational head kinematics and brain tissue strain from a 2-D parametric finite element analysis, Brain Multiphysics. 2 (2021) 100024. https://doi.org/10.1016/j.brain.2021.100024.

[34]  R.M. Wright, A. Post, B. Hoshizaki, K.T. Ramesh, A Multiscale Computational Approach to Estimating Axonal Damage under Inertial Loading of the Head, J. Neurotrauma. 30 (2013) 102–118. https://doi.org/10.1089/neu.2012.2418.

[35]  A. Alshareef, A.K. Knutsen, C.L. Johnson, A. Carass, K. Upadhyay, P. V Bayly, D.L. Pham, J.L. Prince, K.T. Ramesh, Integrating Material Properties from Magnetic Resonance Elastography into Subject-Specific Computational Models for the Human Brain, Brain Multiphysics. (2021) 100038. https://doi.org/10.1016/j.brain.2021.100038.

[36]  S. Wakana, L.M. Nagae-Poetscher, H. Jiang, P. van Zijl, X. Golay, S. Mori, Macroscopic orientation component analysis of brain white matter and thalamus based on diffusion tensor imaging, Magn. Reson. Med. 53 (2005) 649–657. https://doi.org/10.1002/mrm.20386.

[37]  C. Giordano, S. Kleiven, Evaluation of axonal strain as a predictor for mild traumatic brain injuries using finite element modeling, Stapp Car Crash J. 58 (2014) 29–61. https://www.sae.org/content/2014-22-0002/.

[38]  K. Upadhyay, G. Subhash, D. Spearot, Visco-hyperelastic constitutive modeling of strain rate sensitive soft materials, J. Mech. Phys. Solids. 135 (2020) 103777. https://doi.org/10.1016/j.jmps.2019.103777.

[39]  X. Jin, F. Zhu, H. Mao, M. Shen, K.H. Yang, A comprehensive experimental study on material properties of human brain tissue, J. Biomech. 46 (2013) 2795–2801. https://doi.org/10.1016/j.jbiomech.2013.09.001.

[40]  Z. Li, C. Ji, D. Li, R. Luo, G. Wang, J. Jiang, A comprehensive study on the mechanical properties of different regions of 8-week-old pediatric porcine brain under tension, shear, and compression at various strain rates, J. Biomech. 98 (2020) 109380. https://doi.org/10.1016/j.jbiomech.2019.109380.

[41]  H. Mao, L. Zhang, B. Jiang, V. V. Genthikatti, X. Jin, F. Zhu, R. Makwana, A. Gill, G. Jandir, A. Singh, K.H. Yang, Development of a Finite Element Human Head Model Partially Validated With Thirty Five Experimental Cases, J. Biomech. Eng. 135 (2013). https://doi.org/10.1115/1.4025101.

[42]  Y.-C. Lu, N.P. Daphalapurkar, A.K. Knutsen, J. Glaister, D.L. Pham, J.A. Butman, J.L. Prince, P. V. Bayly, K.T. Ramesh, A 3D Computational Head Model Under Dynamic Head Rotation and Head Extension Validated Using Live Human Brain Data, Including the Falx and the Tentorium, Ann. Biomed. Eng. 47 (2019) 1923–1940. https://doi.org/10.1007/s10439-019-02226-z.

[43]  J.E. Galford, J.H. McElhaney, A viscoelastic study of scalp, brain, and dura, J. Biomech. 3 (1970) 211–221. https://doi.org/10.1016/0021-9290(70)90007-2.

[44]  J.H. McElhaney, J.W. Melvin, V.L. Roberts, H.D. Portnoy, Dynamic Characteristics of the Tissues of the Head, in: R.M. Kenedi (Ed.), Perspect. Biomed. Eng., Palgrave Macmillan UK, London, 1973: pp. 215–222. https://doi.org/10.1007/978-1-349-01604-4_34.

[45]  W. Goldsmith, Biomechanics of head injury, in: Y. Fung (Ed.), Biomech. Its Found. Object.,

Prentice Hall, Englewood Cliffs, NJ, 1972: pp. 585–634.

[46] J.R. MACDONALD, Some Simple Isothermal Equations of State, Rev. Mod. Phys. 38 (1966) 669–679. https://doi.org/10.1103/RevModPhys.38.669.

[47] M.C. Kennedy, A. O'Hagan, Bayesian calibration of computer models, J. R. Stat. Soc. Ser. B (Statistical Methodol. 63 (2001) 425–464. https://doi.org/10.1111/1467-9868.00294.

[48] K. Kontolati, D. Loukrezis, K.R.M. dos Santos, D.G. Giovanis, M.D. Shields, Manifold learning-based polynomial chaos expansions for high-dimensional surrogate models, (2021) 1–29. http://arxiv.org/abs/2107.09814.

[49] T. O'Leary-Roseberry, U. Villa, P. Chen, O. Ghattas, Derivative-informed projected neural networks for high-dimensional parametric maps governed by PDEs, Comput. Methods Appl. Mech. Eng. 388 (2022) 114199. https://doi.org/10.1016/j.cma.2021.114199.

[50] D.G. Giovanis, M.D. Shields, Data-driven surrogates for high dimensional models using Gaussian process regression on the Grassmann manifold, Comput. Methods Appl. Mech. Eng. 370 (2020) 113269. https://doi.org/10.1016/j.cma.2020.113269.

[51] B. Bah, Diffusion Maps : Analysis and Applications, University of Oxford, 2008.

[52] K.R.M. dos Santos, D.G. Giovanis, K. Kontolati, D. Loukrezis, M.D. Shields, Grassmannian diffusion maps based surrogate modeling via geometric harmonics, (2021) 1–27. http://arxiv.org/abs/2109.13805.

[53] A. Olivier, D.G. Giovanis, B.S. Aakash, M. Chauhan, L. Vandanapu, M.D. Shields, UQpy: A general purpose Python package and development environment for uncertainty quantification, J. Comput. Sci. 47 (2020) 101204. https://doi.org/10.1016/j.jocs.2020.101204.

[54] D. Lehmberg, F. Dietrich, G. Köster, H.-J. Bungartz, Datafold: Data-Driven Models for Point Clouds and Time Series on Manifolds, J. Open Source Softw. 5 (2020) 2283. https://doi.org/10.21105/joss.02283.

[55] S.E. Davis, S. Cremaschi, M.R. Eden, Efficient Surrogate Model Development: Impact of Sample Size and Underlying Model Dimensions, in: Comput. Aided Chem. Eng., 2018: pp. 979–984. https://doi.org/10.1016/B978-0-444-64241-7.50158-0.

[56] F. George, B.M. Golam Kibria, Confidence intervals for estimating the population signal-to-noise ratio: a simulation study, J. Appl. Stat. 39 (2012) 1225–1240. https://doi.org/10.1080/02664763.2011.644527.

[57] L.F. Gabler, J.R. Crandall, M.B. Panzer, Development of a Metric for Predicting Brain Strain Responses Using Head Kinematics, Ann. Biomed. Eng. 46 (2018) 972–985. https://doi.org/10.1007/s10439-018-2015-9.

[58] L. V Hiscox, H. Schwarb, M.D.J. McGarry, C.L. Johnson, Aging brain mechanics: Progress and promise of magnetic resonance elastography, Neuroimage. 232 (2021) 117889. https://doi.org/10.1016/j.neuroimage.2021.117889.

[59]    L. V. Hiscox, M.D.J. McGarry, H. Schwarb, E.E.W. Van Houten, R.T. Pohlig, N. Roberts, G.R. Huesmann, A.Z. Burzynska, B.P. Sutton, C.H. Hillman, A.F. Kramer, N.J. Cohen, A.K. Barbey, K.D. Paulsen, C.L. Johnson, Standard-space atlas of the viscoelastic properties of the human brain, Hum. Brain Mapp. 41 (2020) 5282–5300. https://doi.org/10.1002/hbm.25192.

[60]    S. Kleiven, Predictors for traumatic brain injuries evaluated through accident reconstructions., Stapp Car Crash J. 51 (2007) 81–114. https://www.sae.org/content/2007-22-0003/.